\documentclass{aa}  

\usepackage{graphicx}
\usepackage{subcaption}
\usepackage{txfonts}
\usepackage{hyperref}

\usepackage{suffix}
\usepackage[printonlyused]{acro}

\acsetup{
  single        = false,
  list/sort     = true,
  cite/group    = true,
  cite/group/cmd = \citealp,
  cite/group/pre = {; },
}

\DeclareAcronym{EVE}{
  short = EVE,
  long = \textit{Extreme Ultraviolet Variability Experiment},
  cite = {Woods:2012}}

\DeclareAcronym{AIA}{
  short = AIA,
  long = \textit{Atmospheric Imaging Assembly},
  cite = {Lemen:2012}}

\DeclareAcronym{HMI}{
  short = HMI,
  long = \textit{Helioseismic Magnetic Imager},
  cite = {Scherrer:2012}}
  
\DeclareAcronym{SDO}{
  short = SDO,
  long = \textit{Solar Dynamics Observatory},
  cite = {Pesnell:2012}}

\DeclareAcronym{STEREO}{
  short = STEREO,
  long = \textit{Solar Terrestrial Relations Observatory},
  cite = {Kaiser:2008}}

\DeclareAcronym{SECCHI}{
  short = SECCHI,
  long = \textit{Sun Earth Connection Coronal and Heliospheric Investigation},
  cite = {Howard:2008}}
  
\DeclareAcronym{SCIP}{
  short = SCIP,
  long = Sun Centered Imaging Package}
  
\DeclareAcronym{MGN}{
  short = MGN,
  long = multi-scale Gaussian normalisation,
  cite = {Morgan:2014}}

\DeclareAcronym{EUVI}{
  short = EUVI,
  long = EUVI}

\DeclareAcronym{CME}{
  short = CME,
  short-plural-form = CMEs,
  long = coronal mass ejection,
  long-plural-form = coronal mass ejections}
  
\DeclareAcronym{PIL}{
  short = PIL,
  short-plural-form = PILs,
  long = polarity inversion line,
  long-plural-form = polarity inversion lines}
  
\DeclareAcronym{LOS}{
  short = LOS,
  short-plural-form = LOS,
  long = line of sight,
  long-plural-form = lines of sight}
  
\DeclareAcronym{FOV}{
  short = FOV,
  short-plural-form = FOV,
  long = field of view,
  long-plural-form = fields of view}
  
\DeclareAcronym{UV}{
  short = UV,
  short-plural-form = UV,
  long = ultraviolet,
  long-plural-form = ultraviolet}

\DeclareAcronym{EUV}{
  short = EUV,
  short-plural-form = EUV,
  long = extreme ultraviolet,
  long-plural-form = extreme ultraviolet}
  
\DeclareAcronym{GONG}{
  short = GONG,
  long = \textit{Global Oscillations Network Group}}
  
\DeclareAcronym{DST}{
  short = DST,
  long = \textit{Dunn Solar Telescope}}

\DeclareAcronym{IBIS}{
  short = IBIS,
  long = \textit{Interferometric Bidimensional Spectropolarimeter},
  cite = {Cavallini:2006,Reardon:2008}}

\DeclareAcronym{FIRS}{
  short = FIRS,
  long = \textit{Facility Infrared Spectropolarimeter},
  cite = {Jaeggli:2011}}
  
\DeclareAcronym{EIT}{
  short = EIT,
  long = \textit{Extreme Ultraviolet Imaging Telescope},
  cite = {Delaboudiniere:1995}}
 
\DeclareAcronym{ROSA}{
  short = ROSA,
  long = \textit{Rapid Oscillations in the Solar Atmosphere},
  cite = {Jess:2010}}
  
\DeclareAcronym{SPINOR}{
  short = SPINOR,
  long = \textit{Spectro-Polarimeter for Infrared and Optical Regions},
  cite = {Socas:2006}}  
 
\DeclareAcronym{SOHO}{
  short = SOHO,
  long = \textit{Solar and Heliospheric Observatory},
  cite = {Domingo:1995}}
 
\DeclareAcronym{MDI}{
  short = MDI,
  long = \textit{Michelson Doppler Imager},
  cite = {Scherrer:1995}}

\DeclareAcronym{NMSU}{
  short = NMSU,
  long = New Mexico State University}

\DeclareAcronym{NSO}{
  short = NSO,
  long = \textit{National Solar Observatory}}
  
\DeclareAcronym{TAC}{
  short = TAC,
  long = time allocation committee}
  
\DeclareAcronym{DEM}{
  short = DEM,
  short-plural-form = DEMs,
  long = Differential Emission Measure,
  long-plural-form = Differential Emission Measures}
  
\DeclareAcronym{PI}{
  short = PI,
  short-plural-form = PI,
  long = Principle Investigator,
  long-plural-form = Principle Investigators}  
  
\DeclareAcronym{IDL}{
  short = IDL,
  long = Interactive Data Language}
  
\DeclareAcronym{bb}{
  short = bb,
  long = broadband}  
  
\DeclareAcronym{nb}{
  short = nb,
  long = narrowband}    
  
\DeclareAcronym{IPM}{
  short = IPM,
  long = interplanetary medium}  
  
\DeclareAcronym{AO}{
  short = AO,
  long = adaptive optics,
  cite = {Rimmele:2004}} 
  
\DeclareAcronym{IR}{
  short = IR,
  long = infrared}  

\DeclareAcronym{halpha}{
  short = H-$\alpha$,
  long = Hydrogen-$\alpha$}

\DeclareAcronym{FPI}{
  short = FPI,
  long = Fabry-P\'erot}
  
\DeclareAcronym{DWDM}{
  short = DWDM,
  long = dense wavelength division multiplexing}  

\DeclareAcronym{CCD}{
  short = CCD,
  long = charge-coupled device}   
  
\DeclareAcronym{HI}{
  short = HI,
  long = Heliospheric Investigation}   
  
\DeclareAcronym{ROI}{
  short = ROI,
  long = region-of-interest}    
  
\DeclareAcronym{POV}{
  short = POV,
  long = point-of-view}  
  
\DeclareAcronym{MHS}{
  short = MHS,
  long = magnetohydrostatic}
  
\DeclareAcronym{MHD}{
  short = MHD,
  long = magnetohydrodynamic}  
  
\DeclareAcronym{RMHD}{
  short = RMHD,
  long = radiative magnetohydrodynamic}  
  
\DeclareAcronym{DOT}{
  short = DOT,
  long = \textit{Dutch Open Telescope},
  cite = {Rutten:1997}}  
  
\DeclareAcronym{BBSO}{
  short = BBSO,
  long = \textit{Big Bear Solar Observatory}}
  
\DeclareAcronym{NST}{
  short = NST,
  long = \textit{New Solar Telescope},
  cite = {Goode:2012}}

\DeclareAcronym{SOT}{
  short = SOT,
  long = \textit{Solar Optical Telescope},
  cite = {Tsuneta:2008}}
  
\DeclareAcronym{hinode}{
  short = Hinode,
  long = Hinode,
  cite = {Kosugi:2007}}
  
\DeclareAcronym{GST}{
  short = GST,
  long = \textit{Goode Solar Telescope}}  
  
\DeclareAcronym{SST}{
  short = SST,
  long = \textit{Swedish 1-m Solar Telescope},
  cite = {Scharmer:2002}}  
  
\DeclareAcronym{TRACE}{
  short = TRACE,
  long = \textit{Transition Region and Coronal Explorer},
  cite = {Handy:1999}}  
  
\DeclareAcronym{PCTR}{
  short = PCTR,
  long = \textit{prominence-corona-transition-region}}
  
\DeclareAcronym{DKIST}{
  short = DKIST,
  long = Daniel K. Inouye Solar Telescope}
  
\DeclareAcronym{RTI}{
  short = RTI,
  long = Rayleigh-Taylor instability}
  
\DeclareAcronym{SSW}{
  short = SSW,
  long = SolarSoftWare,
  cite = {Freeland:1998}}    
  
\DeclareAcronym{LTE}{
  short = LTE,
  long = local thermodynamic equilibrium}
  
\DeclareAcronym{NLTE}{
  short = NLTE,
  long = non-"local thermodynamic equilibrium"}
  
\DeclareAcronym{FTS}{
  short = FTS,
  long = Fourier Transform Spectrometer,
  cite = {Kurucz:1984}}
  
\DeclareAcronym{RTE}{
  short = RTE,
  long = radiative transfer equation}
  
\DeclareAcronym{CRD}{
  short = CRD,
  long = complete frequency redistribution}
  
\DeclareAcronym{PRD}{
  short = PRD,
  long = partial frequency redistribution}
  
\DeclareAcronym{BCM}{
	short = BCM,
	long = Beckers' cloud model,
	cite={Beckers:1964}}
	
\DeclareAcronym{HSRA}{
	short = HSRA,
	long = Harvard Smithsonian Reference Atmosphere,
	cite={Gingerich:1971}}
	
\DeclareAcronym{NICOLE}{
	short = NICOLE,
	long = Non-LTE Inversion COde using the Lorien Engine,
	cite={Socasnavarro:2015}}
	
\DeclareAcronym{SIR}{
	short = SIR,
	long = Stokes Inversion based on Response functions,
	cite={Ruizcobo:1992}}
	
\DeclareAcronym{HAZEL}{
	short = HAZEL,
	long = Hanle and Zeeman Light,
	cite={AsensioRamos:2008}}
	
\DeclareAcronym{BPSS}{
	short = BPSS,
	long = bald patch separatrix surface,
	cite={Bungey:1996}}
	
\DeclareAcronym{EIS}{
	short = EIS,
	long = \textit{EUV Imaging Spectrometer},
	cite={Culhane:2007}}

\DeclareAcronym{FWHM}{
  short = FWHM,
  long = full width at half maximum}

\DeclareAcronym{AMRVAC}{
	short = MPI-AMRVAC,
	long = \textit{Adaptive Mesh Refinement Versatile Advection Code},
	cite = {Keppens:2012,Porth:2014,Xia:2018,Keppens:2020}}

\DeclareAcronym{AMR}{
	short = AMR,
	long = adaptive mesh refinement}

\DeclareAcronym{CCI}{
	short = CCI,
	long = Convective Continuum Instability}

\DeclareAcronym{BV}{
	short = BV,
	long = Brunt-V\"ais\"al\"a}

\DeclareAcronym{TVDLF}{
	short = TVDLF,
	long = Total Variation Diminishing Lax-Friedrich}

\DeclareAcronym{TI}{
	short = TI,
	long = Thermal Instability}

\DeclareAcronym{TNE}{
	short = TNE,
	long = thermal non-equilibrium}
	
\DeclareAcronym{MALI}{
    short = MALI,
    long = Multilevel Accelerated Lambda Iteration}
    
\DeclareAcronym{yt}{
    short = yt,
    long = yt-project,
    cite = {Turk:2011}}
    
\DeclareAcronym{GL}{
    short = GL,
    long = Gauss-Legendre}
    
\DeclareAcronym{CM}{
    short = CM,
    long = classically mounted}
    
\DeclareAcronym{MULTI3D}{
    short = MULTI3D,
    long = \textit{multi-level non-LTE 3D},
    cite = {Leenaarts:2009}}

\DeclareAcronym{EB}{
    short = EB,
    long = \textit{Eddington-Barbier}}
    
\DeclareAcronym{KDE}{
    short = KDE,
    long = kernel density estimate}
    
\DeclareAcronym{NLFFF}{
    short = NLFFF,
    long = nonlinear force-free field}

\usepackage[normalem]{ulem}
\usepackage[x11names]{xcolor}
\usepackage{academicons}
\definecolor{orcidlogocol}{HTML}{A6CE39}

\newcommand{\orcid}[1]{\href{https://orcid.org/#1}{\textcolor[HTML]{A6CE39}{\aiOrcid}}}
\usepackage{xcolor}
\usepackage{soul}

\begin{document}

   \title{3D coupled tearing-thermal evolution in solar current sheets}


   \author{Samrat Sen \href{https://orcid.org/0000-0003-1546-381X}{\includegraphics[scale=0.05]{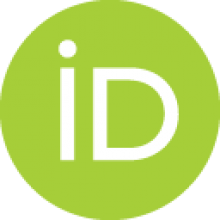}}
          \inst{1}$^*$, Jack Jenkins\href{http://orcid.org/0000-0002-8975-812X} {\includegraphics[scale=0.05]{orcid-ID.png}} \inst{1},
          \and
          Rony Keppens\href{https://orcid.org/0000-0003-3544-2733}{\includegraphics[scale=0.05]{orcid-ID.png}} \inst{1}
          }

   \institute{$^1$Centre for mathematical Plasma-Astrophysics, Celestijnenlaan 200B, 3001 Leuven, KU Leuven, Belgium\\
              $^*$\email{samratseniitmadras@gmail.com}
                      }

   \date{Received: XXXX; accepted: XXXX}
   \date{}
 
  \abstract
   {The tearing instability plays a major role in the disruption of current sheets, whereas thermal modes can be responsible for condensation phenomena (forming prominences and coronal rain) in the solar atmosphere. However, how combined tearing-thermal unstable current sheets evolve within the solar atmosphere has received limited attention to date.}
   {We numerically explore a combined tearing and thermal instability that causes the break up of an idealized current sheet in the solar atmosphere. The thermal component leads to the formation of localized, cool condensations within an otherwise 3D reconnecting magnetic topology.}
   {We construct a 3D resistive magnetohydrodynamic simulation of a force-free current sheet under solar atmospheric conditions that incorporate the non-adiabatic influence of background heating, optically thin radiative energy loss, and magnetic field aligned thermal conduction with the open source code \texttt{MPI-AMRVAC}. Multiple levels of adaptive mesh refinement reveal the self-consistent development of finer-scale condensation structures within the evolving system.}
   {The instability in the current sheet is triggered by magnetic field perturbations concentrated around the current sheet plane, and subsequent tearing modes develop. This in turn drives thermal runaway associated with the thermal instability of the system. We find subsequent, localized cool plasma condensations that form under the prevailing low plasma-$\beta$ conditions, and demonstrate that the density and temperature of these condensed structures are similar to more quiescent coronal condensations. Synthetic counterparts at Extreme-UltraViolet (EUV) and optical wavelengths show the formation of plasmoids (in EUV), and coronal condensations similar to prominences and coronal rain blobs in the vicinity of the reconnecting sheet.}
   {Our simulations imply that 3D reconnection in solar current sheets may well present an almost unavoidable multi-thermal aspect, that forms during their coupled tearing-thermal evolution.}


\titlerunning{
Coupled tearing-thermal evolutions}
\authorrunning{Sen et al.}

\maketitle


\section{Introduction} 

Magnetic reconnection is a fundamental process understood to play a critical role throughout the solar atmosphere. The change of magnetic field topology during reconnection leads to conversion of magnetic energy into thermal and kinetic energies \citep{2000mrp..book.....B}, frequently leading to fast energy release of solar flares \citep{1939ApJ....89..555G, 1947MNRAS.107..338G, 1948MNRAS.108..163G, 2000mare.book.....P, 2020JGRA..12525935H} and coronal mass ejections \citep{1995GeoRL..22.1753G, 2003JGRA..108.1023S, 2012ApJ...760...81K} out into the heliosphere.

It was suggested by \cite{1963PhFl....6..459F} that reconnection in an incompressible plasma may be triggered due to small perturbations in a current layer, correspondingly breaking up the current sheet in the form of the tearing instability. Linear analysis by \cite{2007PhPl...14j0703L} suggests that the tearing instability in a single current sheet may lead to the formation of a chain of plasmoids (secondary magnetic islands). This was later verified in 2D numerical simulations by \cite{2013PhPl...20e5702H} and \cite{2013PhPl...20h2131H}. Extensions to 2D double current layer models were seen to give rise to the development and layer-layer interactions of tearing modes with smaller scale plasmoid formation \citep[and references therein]{2011PhPl...18e2303Z, 2013PhPl...20i2109K, 2017PhPl...24h2116A, 2021PhPl...28h2903P}.

Each of these previous studies does not consider non-adiabatic effects of background heating, radiative energy loss, and thermal conduction, which are essential components of the solar atmosphere. It is well established that the solar corona is in an overall delicate thermal balance. If this balance due to optically thin radiative loss, background heating in combination with thermal conduction is perturbed, an increment of the thermal energy loss cooling down the plasma may lead to an enhancement of the plasma density. This in turn radiates more energy (radiative loss in an optically thin medium is proportional to the square of plasma density), and the material becomes cooler still. Hence, a catastrophic runaway process ensues leading to a rapid rise in the density and drop in temperature which is in essence the thermal instability. A detailed linear analysis of the thermal instability is presented by \cite{1953ApJ...117..431P} and \cite{1965ApJ...142..531F}, who derived the criteria governing the onset to a catastrophic radiative loss in an infinite homogeneous medium. The linear magnetohydrodynamic (MHD) analysis was extended to a 1D slab configuration \citep{1991SoPh..131...79V, 1992SoPh..140..317V}, and cylindrical geometry \citep{1991SoPh..134..247V, 2011ApJ...731...39S} under solar coronal conditions. Linear and follow-up nonlinear theory of thermal instability is a powerful tool to explain various fascinating features of the solar atmosphere. For example, the possible formation of a prominence in a current sheet is discussed by \cite{1977SoPh...53...25S} and the dynamic thermal balance in a coronal arcade is studied in \cite{1979SoPh...64..267P}. The post-flare loop formation in a line-tied current sheet configuration with radiative energy loss was simulated in a 2D MHD setup by \cite{1991SoPh..135..361F}. 

More recently, multidimensional simulations related to prominence formation emerged in a variety of magnetic topologies. \cite{2012ApJ...748L..26X} reported the ab initio formation of a solar prominence in a 2.5D MHD simulation in a bipolar magnetic arcade due to chromospheric evaporation and thermal instability. This was revisited in a quadrupolar arcade setup, in which reconnection was induced by the condensing prominence by \cite{2014ApJ...789...22K}. That prominences can also form by feeding chromospheric matter within plasmoids during a flux rope eruption is developed by \cite{2022ApJ...928...45Z}. 3D models of prominence formation that establish the needed plasma cycle between chromosphere and corona are shown in \cite{2016ApJ...823...22X}. More recently, prominence formation due to levitation-condensation \citep{2015ApJ...806..115K, Jenkins:2021} was demonstrated where a 3D realization is needed to allow magnetic Rayleigh-Taylor instability \citep{Jenkins:2022}. The effect of thermal instability has also been explored for the formation and dynamics of coronal rain in magnetic arcades in 2.5D geometry \citep{2013:fan, 2015:fan}, in a weak magnetic bipole in 3D geometry \citep{2017A&A...603A..42X}, in a more self-consistent 3D radiative-magnetohydrodynamic setup in \cite{2020A&A...639A..20K}, and for randomly heated arcades by \cite{2022ApJ...926..216L} in a 2.5D geometry.

These works also triggered a renewed interest in more idealized studies of linear thermal instability and its nonlinear evolution, and in how the various linear MHD waves and instabilities may interact. Numerical analysis in the linear and non-linear domains due to the interaction of the slow MHD and entropy (thermal) modes was carried out in recent studies by \cite{2019A&A...624A..96C} and \cite{2020A&A...636A.112C}, while the effect of different radiative loss functions on the onset and far nonlinear behavior of thermal modes was analyzed by \cite{2021A&A...655A..36H}. However, the influence of thermal instability on the tearing mode of solar current sheets has not gained much attention to date. Linear analysis by \cite{2021SoPh..296...74L, 2021SoPh..296...93L, 2021SoPh..296..117L} shows that the instability growth rate in a pre-flare current sheet is modified if the non-adiabatic effects of radiative energy loss, resistivity and thermal conductivities are included. \cite{2022A&A...666A..28S}[SK22 hereafter] extended this into the non-linear domain and incorporated background heating and optically thin radiative loss into a series of 2D resistive MHD simulations. This study finds that the instability growth rate of tearing modes in a solar current sheet increases by an order of magnitude when these non-adiabatic effects are incorporated, such that we can meaningfully speak of coupled tearing-thermal evolutions. The 2D current sheet produced a chain of plasmoid-trapped condensations with cool material, which are thermodynamically similar to prominence (or coronal rain) in the solar atmosphere. 

In this work, we extend our study to a 3D geometry and explore the tearing-thermal evolutionary process of an idealized current sheet model in solar atmosphere which is essentially non-adiabatic (with background heating, optically-thin radiative loss, and thermal conduction). Due to the mutual reinforcement of these instabilities, we demonstrate the combined effect of the complex evolution of the current layer, which disintegrates into finer structures with subsequent development of flux ropes, along with the formation of cool plasma condensations in the vicinity of this evolving current sheet. These localized cool, and plasma-condensed regions share similarities with the prominence and coronal rain structures observed in the solar atmosphere. Our findings here augment the growing theoretical basis for the combined effect of a current sheet fragmentation, and formation of cool-condensed plasma due to coupled tearing-thermal instability. This multi-mode evolution of the system occurring in association with or during reconnection in a current sheet is an important aspect to understand the dynamics and multi-thermal processes in the solar atmosphere.

The paper is organized as follows. In Sect. \ref{numerical_setup}, we describe the numerical model with an initial configuration with a precise magnetic and thermodynamic structure, and detail algorithmic aspects and boundary conditions. In Sect. \ref{result}, we discuss the main results of the study, and relevance of the model in the solar atmosphere. Section \ref{summary_conclusions} addresses the significance and novelty of the work for a typical coronal atmosphere, the scope for further development and points out how this work will be useful for future studies.



\section{Numerical setup} \label{numerical_setup}
 We construct a resistive MHD simulation using MPI-parallelised Adaptive Mesh Refinement Versatile Advection Code or \texttt{MPI-AMRVAC}\footnote{Open source at: \href{http://amrvac.org}{http://amrvac.org/}} \citep{2012JCoPh.231..718K, 2014ApJS..214....4P, 2018ApJS..234...30X, keppens2021, keppens2023} in 3D Cartesian geometry. The spatial domain of the simulation box spans from $-10$ to $10$ (in dimensionless units) along $x-y-z$ directions. We activate the adaptive mesh refinement (AMR) up to level three, which gives a maximum resolution of $512^3$. If the box size is set in units of 10 Mm, this achieves the smallest cell size of 390 km in each direction. Automated refinement and derefinement is triggered based on the errors estimated by the instantaneous density (gradient) at each time step. 

To explore the influence of thermal instability on the tearing mode in a current sheet, the following normalized MHD equations are solved numerically,
\begin{align} \label{eq:mhd1}
   & \frac{\partial \rho}{\partial t} + \nabla \cdot ({\bf v}\rho) = 0, \\ \label{eq:mhd2}
   & \frac{\partial (\rho {\bf v})}{\partial t}+ \nabla \cdot (\rho {\bf v} {\bf v}+p_{tot}{\bf I}-{\bf B}{\bf B})=\mathbf{0},\\ \label{eq:mhd3}
   & \frac{\partial \mathcal{E}}{\partial t} + \nabla \cdot (\mathcal{E}{\bf v}+p_{tot}{\bf v}-{\bf B}{\bf B}\cdot {\bf v})= \eta \textbf{J}^2 - {\bf B} \cdot \nabla \times (\eta {\bf J})\\\nonumber 
   & \hspace*{2cm} - \rho^2 \Lambda(T)+H_{bgr}+ \nabla \cdot (\kappa_{||} \cdot \nabla T),\\ \label{eq:mhd4}
   & \frac{\partial {\bf B}}{\partial t} + \nabla \cdot ({\bf v}{\bf B}-{\bf B}{\bf v}) + \nabla \times (\eta {\bf J}) = \mathbf{0}\,,\\ \label{eq:mhd5}
   & \nabla \cdot \textbf{B}=0 \,,\\ \label{eq:mhd6}
   & \textbf{J}=\nabla \times \textbf{B} \,.
\end{align}
Note that we use magnetic units where the magnetic permeability is unity. Here, $\textbf{I}$ is the unit tensor, and $\rho, T, {\bf B}$, ${\bf v}$, and $\eta$ represent mass density, temperature, magnetic field vector, velocity vector, and resistivity, respectively. A uniform resistivity, $\eta = 0.001$ (or $1.2 \times 10^{14}$ cm$^2$ s$^{-1}$ in physical units) is taken throughout the entire simulation domain. We adopt the Spitzer-type thermal conductivity, $\kappa_{||} = 10^{-6} T^{5/2}$ erg cm$^{-1}$ s$^{-1}$ K$^{-1}$, which is purely aligned along the magnetic field. The total pressure $p_{tot}$ is the sum of the plasma and magnetic pressure given by
\begin{align}\label{eq:ptot}
    p_{tot} = p+\frac{B^2}{2},
\end{align}
where $p$ is the gas pressure linked with the thermodynamic quantities through the ideal gas law. The total energy density is 
\begin{align}\label{eq:energy_density}
    \mathcal{E}=\frac{p}{\gamma -1} + \frac{\rho v^2}{2} + \frac{B^2}{2},
\end{align}
where $\gamma = 5/3$ is the ratio of specific heats for a monoatomic gas (fully ionized hydrogen plasma). We set up a current sheet configuration using the magnetic field components
\begin{align}\label{Bx}
    B_x &= B_0\ \text{tanh}(z/l_s);\\ \label{By}
    B_y &= \sqrt{B_0^2-B_x^2};\\ \label{Bz}
    B_z &= 0,
\end{align}
where $B_0 = 1$ (corresponding to 2 G in physical units) is the magnetic field strength, which is comparable with the observations in the solar corona, where the field strength at the height of 1.05-1.35 solar radius are reported between 1-4 G \citep{2004ApJ...613L.177L, 2019ApJ...881...24K, 2020ScChE..63.2357Y}. The unit plasma density, temperature, and length scales are set as $\Bar{\rho} = 2.34 \times 10^{-15}$ g cm$^{-3}$, $\bar{T} = 10^6$ K, and $\Bar{L}=10^9$ cm, which are relevant for the solar corona. The initial width of the current sheet is set to $l_s = 0.5$ (5 Mm in physical unit), which is comparable with the observed flare current sheet thickness \citep{Li:2018, savage:2010}. The magnetic field configuration given by Eqns. (\ref{Bx}-\ref{Bz}) represents a force-free field, and the polarity reversal of the magnetic field occurs around the $z=0$ plane, where the current sheet is. In line with a fully force-balanced equilibrium of the system, we use isothermal and isobaric conditions as the initial setup of the model.

The third term at the right-hand side (RHS) of Eq. (\ref{eq:mhd3}) represents the radiative cooling in the optically thin corona, where $\Lambda(T)$ is the cooling function developed by \cite{2008ApJ...689..585C} and extended to lower temperatures following \cite{1972ARA&A..10..375D}. The precise temperature dependence of $\Lambda(T)$ was shown in Figure 1 of our previous 2D simulation (SK22). To maintain the initial thermal balance between optically thin radiative loss and background heating, $H_{bgr}$ of the system, we prescribe a uniform, time-independent value,
\begin{align}
    H_{bgr}=\rho_0^2\ \Lambda(T_0).
\end{align}
The motivation of using the above form is that the radiative cooling term at the initial state exactly compensates the background heating term, and the heating/cooling (mis)balance in the system occurs after a long term evolution triggered by the external perturbations (which is magnetic field in this study). Note that, this heating model is similar, although uniform in space unlike to our earlier study in SK22. However, the role of different heating models based on the power-law behaviour of magnetic field strength, and density on the thermal runaway, and condensation processes have been reported by \cite{2022A&A...668A..47B}, which finds that the different heating models can change the evolution and morphology of the condensations. Therefore, how the different heating rates can change the thermal balance of our model can be interesting to study in future. With a homogeneous density $\rho_0 = 0.2$ ($4.68 \times 10^{-16}$ g cm$^{-3}$ in physical units) and isothermal atmosphere $T_0 = 0.5$ (0.5 MK in physical units) as initial condition, we have an initially uniform plasma-$\beta = 0.2$, less than unity as appropriate for solar corona. We use the initial temperature, $T_0=0.5$ MK, which is in the temperature regime where the cooling function we use in this study has a very sharp gradient, and the heating-cooling mis-balance may be dominated due to perturbation from the equilibrium temperature. However, we also notice that the system achieves to thermal runaway state, and cool-condensed materials form for other equilibrium temperature regime, $T_0=1$ MK, which is shown in Appendix \ref{app}. It is to be noted from the last term in the RHS of Eq.~\ref{eq:mhd3}, that there is no role for thermal conduction at the initial time, as the system starts off isothermal. Therefore, the system is initially in thermal equilibrium, but the finite value of resistivity will drive the ideal force-balanced state away from its initial state, but only on the (slow) resistive timescale. This setup is liable to both linear resistive tearing modes, for which finite resistivity is key, and has all thermodynamic ingredients to allow for thermal instability. The equilibrium system is perturbed to trigger tearing modes, which can further trigger the thermal modes, and thus enforce each other in a coupled tearing-thermal fashion. We use parametrically controlled, monopole-free magnetic field perturbations mainly confined in the vicinity of the $z=0$ plane (where the initial current sheet is present) and exponentially decaying for $|z| >0$,
\begin{align}\label{dbx}
    \delta B_x = &-\frac{2\pi}{l}\bigg[\psi_{01}\ \text{cos}\bigg(\frac{2\pi n_1 x}{l}\bigg)\ \text{sin}\bigg(\frac{2\pi n_1 y}{l}\bigg)\\ \nonumber &+ \psi_{02}\ \text{cos}\bigg(\frac{2\pi n_2 x}{l}\bigg)\ \text{sin}\bigg(\frac{2\pi n_2 y}{l}\bigg)\bigg]\ \text{Exp}(-z^2/l_s),\\ \label{dby}
    \delta B_y = &+\frac{2\pi}{l}\bigg[\psi_{01}\ \text{sin}\bigg(\frac{2\pi n_1 x}{l}\bigg)\ \text{cos}\bigg(\frac{2\pi n_1 y}{l}\bigg)\\ \nonumber &+ \psi_{02}\ \text{sin}\bigg(\frac{2\pi n_2 x}{l}\bigg)\ \text{cos}\bigg(\frac{2\pi n_2 y}{l}\bigg)\bigg]\ \text{Exp}(-z^2/l_s).
\end{align}
Here, the parameter $l = 20 \times \Bar{L}$ matches the geometric sizes of the simulation domain along $x$ and $y$ directions respectively, the perturbation amplitudes $\psi_{01} = \psi_{02} = 0.1$ ensure a variation of $10 \%$ of the magnetic field strength $B_0$, and we take the multi-mode distribution of the perturbations using $n_1 =4$ and $n_2 =2$. The magnetic field distribution (Eqs. \ref{Bx}-\ref{Bz}), and the perturbations (Eqs. \ref{dbx}-\ref{dby}) ensure the solenoidal condition, $\nabla \cdot {\bf B} = \nabla \cdot \delta{\bf B} = 0$. 

After the initial setup, the system is allowed to evolve as governed by the Eqs. (\ref{eq:mhd1}-\ref{eq:mhd6}). The equations are solved numerically using a three-step Runge-Kutta time integration with a second order slope limited reconstruction method \citep{2006:ruuth} with `Vanleer' flux limiter \citep{1974:vanLeer}, and a Total Variation Diminishing Lax-Friedrichs (TVDLF) flux scheme. We follow the evolution of the system for up to 214.7 minutes and save the data at a cadence of 85.87 s, which gives 151 snapshots. We use periodic boundary conditions along $x, y$ directions, and open boundary condition along the $z$ direction. The wall clock time for the entire simulation run is $\approx 90$ hours using 8 nodes with 288 processors in total with GNU-Fortran (version 6.4.0) compiler and open MPI 2.1.2.

Note that, there are some important differences in the initial conditions of this model with respect to SK22 mentioned in the following. (i) This is a force-free magnetic field configuration, and therefore an isobaric and isothermal medium ensures a fully force-balanced equilibrium, whereas the magnetic field configuration in SK22 was non-force free, and therefore we used a non-uniform density profile (though the initial temperature was also uniform) in such a way that it maintained the force-balance equilibrium, (ii) The magnetic field strength near the current sheet in SK22 is $\ll 1$, which sets the plasma-$\beta \gg 1$ near the current sheet, on the other hand, our current model has initially a uniform and low plasma-$\beta$ = 0.2 throughout the entire simulation domain, (iii) the direction of imposed magnetic field perturbations for SK22 were both parallel and perpendicular to the current sheet, but in the current work we set the perturbation only parallel and concentrated around the current sheet plane. Besides the difference in affordable numerical resolution (SK22 has the maximum resolution of $2048^2$, whereas the current setup has maximum resolution of $512^3$), the system studied here is intrinsically 3D, and is hence more relevant for actual current sheet conditions.

\section{Result and Discussions} \label{result}

\begin{figure*}[hbt!]
\centering
\begin{subfigure}{0.49\textwidth}
  \includegraphics[width=1\textwidth]{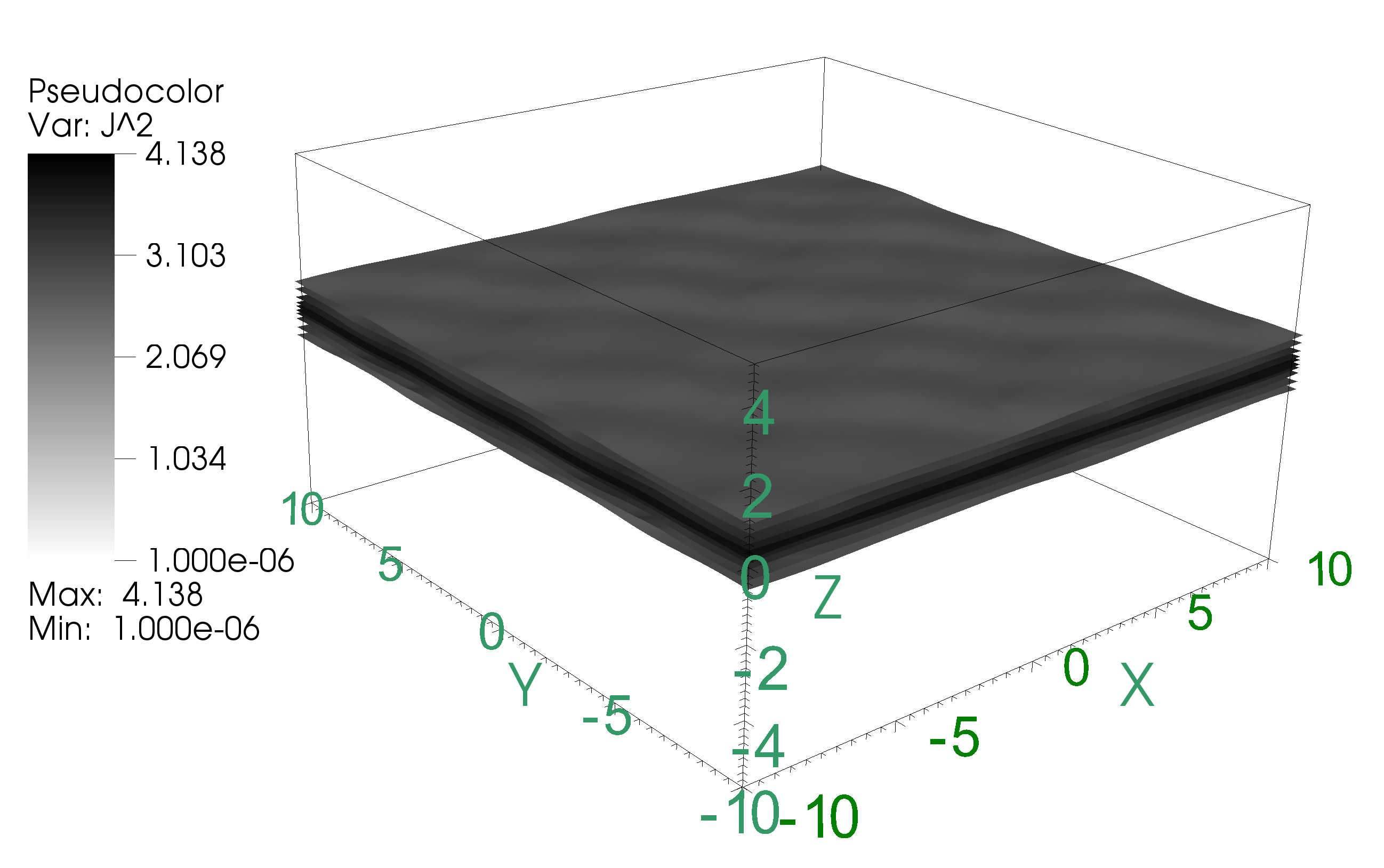}
   \caption{t=14.3 min}
   \label{fig:jsqr_t010}
\end{subfigure}
\begin{subfigure}{0.49\textwidth}
    \includegraphics[width=1\textwidth]{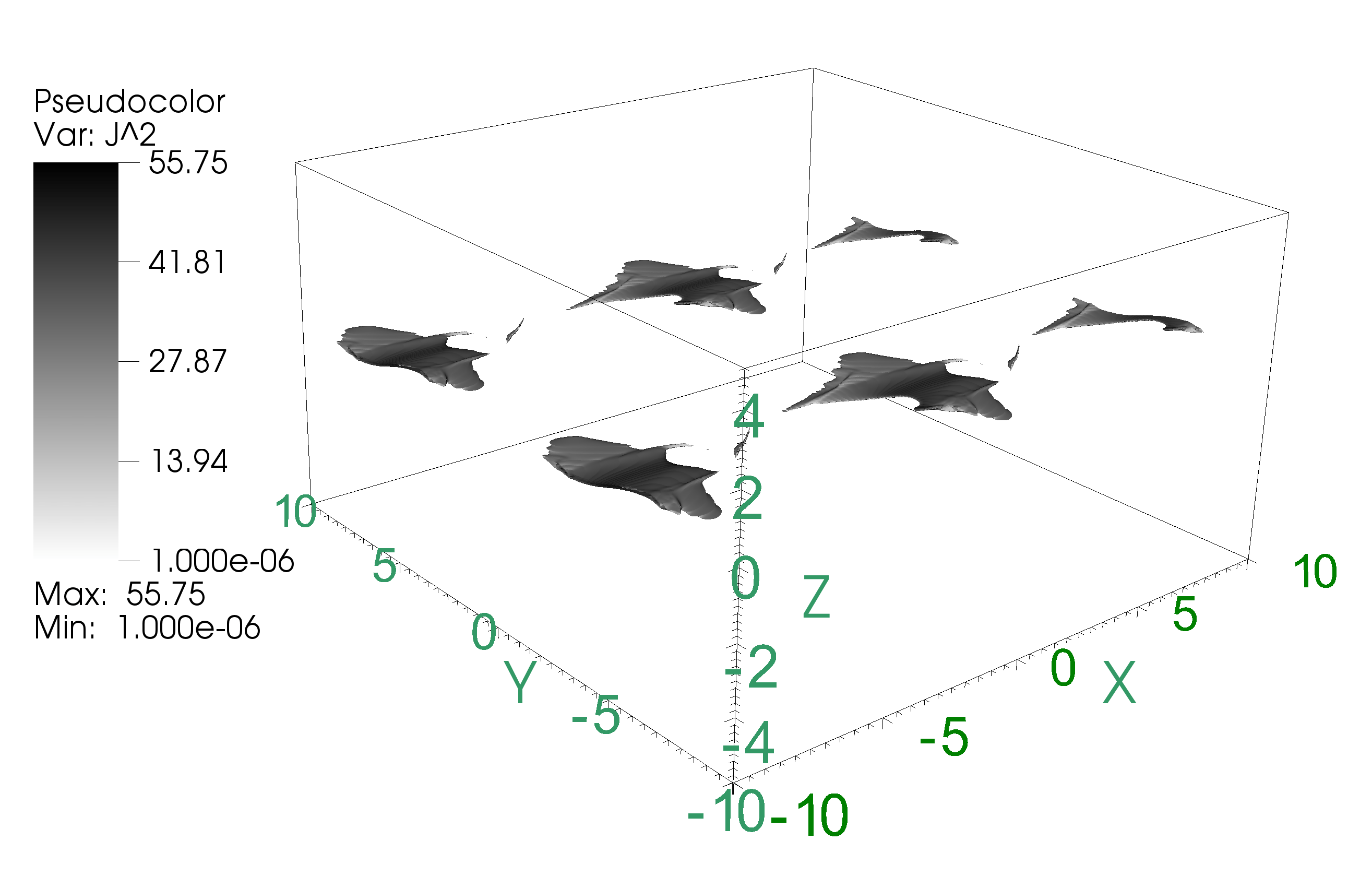}
    \caption{t=207.5 min}
    \label{fig:jsqr_t145}
\end{subfigure}
\newline
\centering
\begin{subfigure}{0.49\textwidth}
  \includegraphics[width=0.8\textwidth]{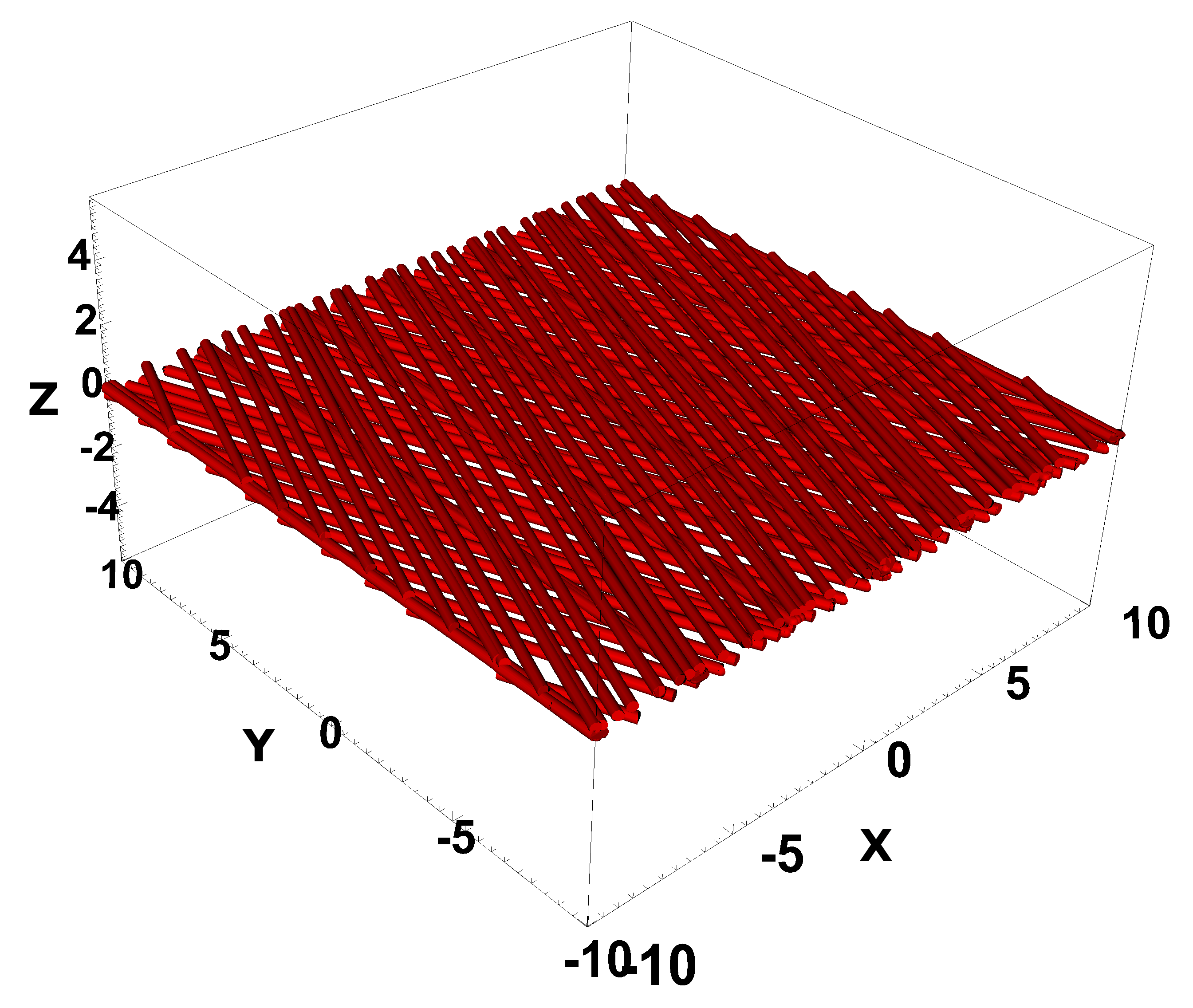}
   \caption{t=14.3 min}
   \label{fig:fl_t003}
\end{subfigure}
\begin{subfigure}{0.49\textwidth}
    \includegraphics[width=0.8\textwidth]{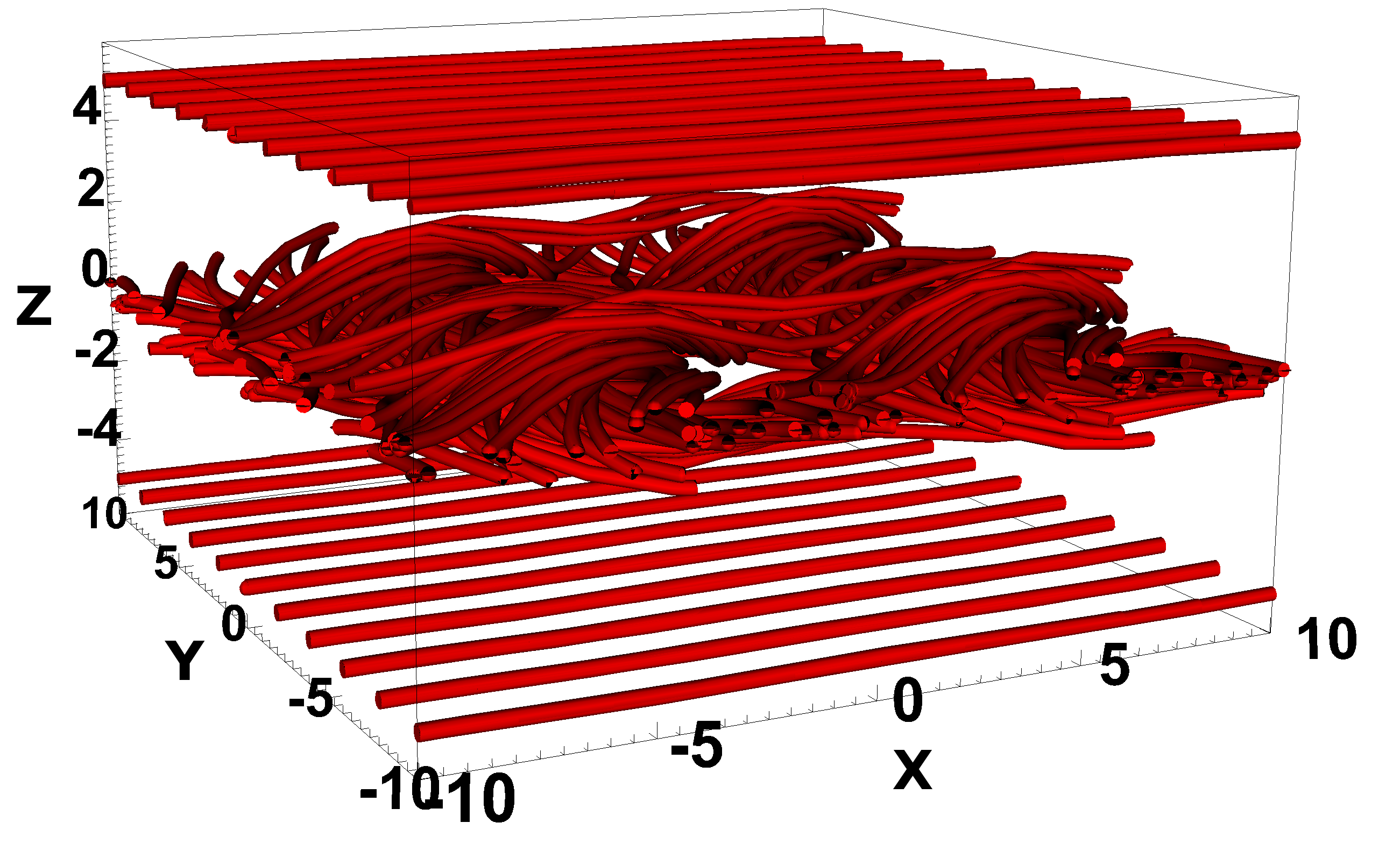}
    \caption{t=207.5 min}
    \label{fig:fl_t145}
\end{subfigure}
\caption{Isosurface (five isosurface values are taken from instantaneous minimum to maximum) plots of the current density squared $J^2$, which represent the 3D structure of the evolving current sheet. Panel (a) represents the perturbations appearing in the current sheet plane due to the multimode ($n_1=4$, $n_2=2$) magnetic field perturbations in Eqns. (\ref{dbx}-\ref{dby}), (b) represents a time when the current sheet disintegrated due to tearing-thermal coupled evolution of the system. Panels (c) and (d) represent the magnetic field corresponding to panels (a) and (b), respectively. Distances along $x, y,$ and $z$ axes are scaled in units of $10^4$ km, where we see the field lines perturbed at 14.3 min around the current sheet plane only, which evolves to sizeable flux rope structures at 207.5 min. (An animation of the figures is available online.)} 
\label{fig:jsqr}
\end{figure*}

\begin{figure*}[hbt!]
\centering
\begin{subfigure}{0.3\textwidth}
    \includegraphics[width=1.1\textwidth]{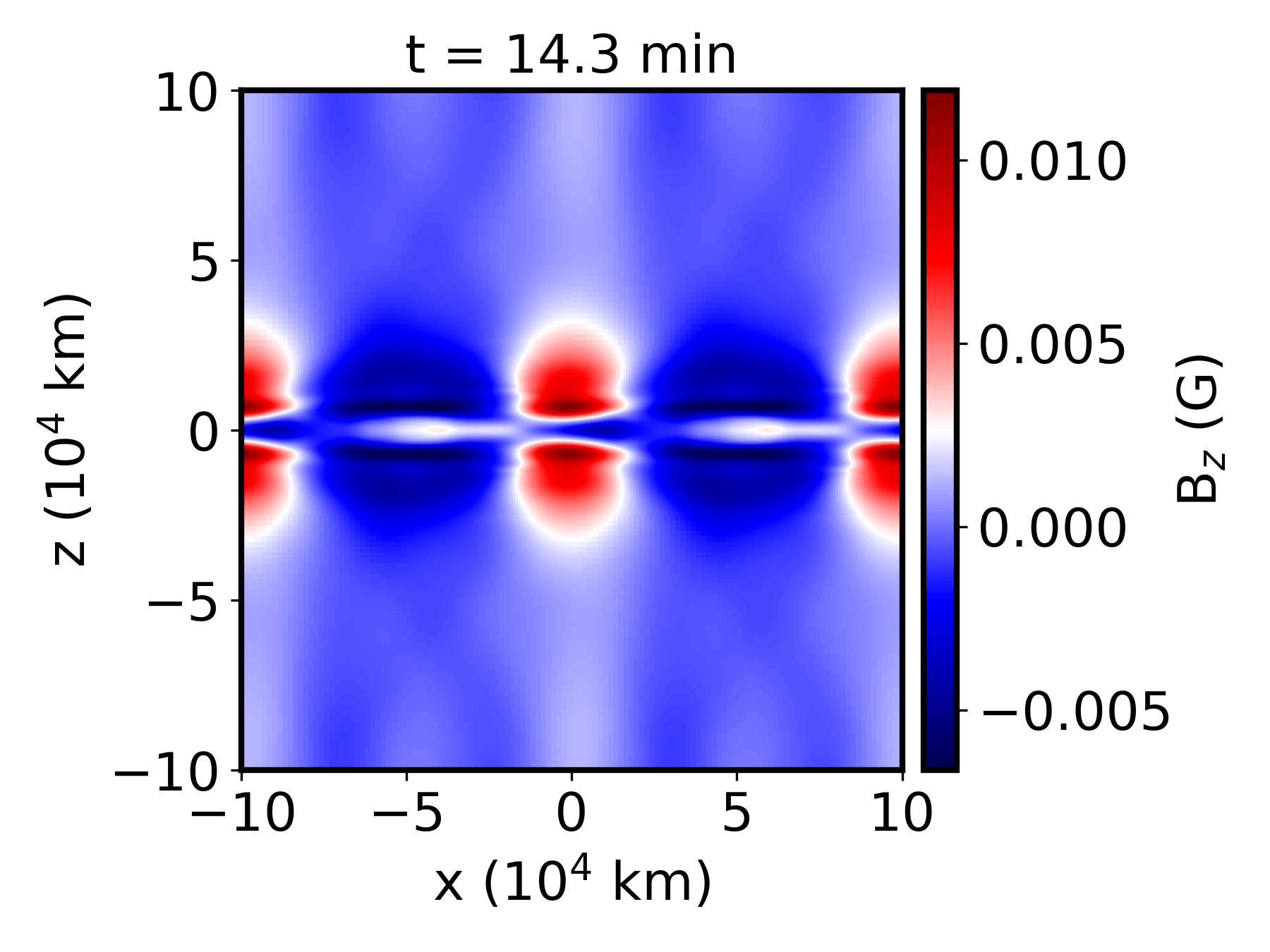}
    \caption{at $y=0$}
    \label{fig:bz_xz_t010}
\end{subfigure}
\hspace{0.5 cm}
\begin{subfigure}{0.3\textwidth}
    \includegraphics[width=1.1\textwidth]{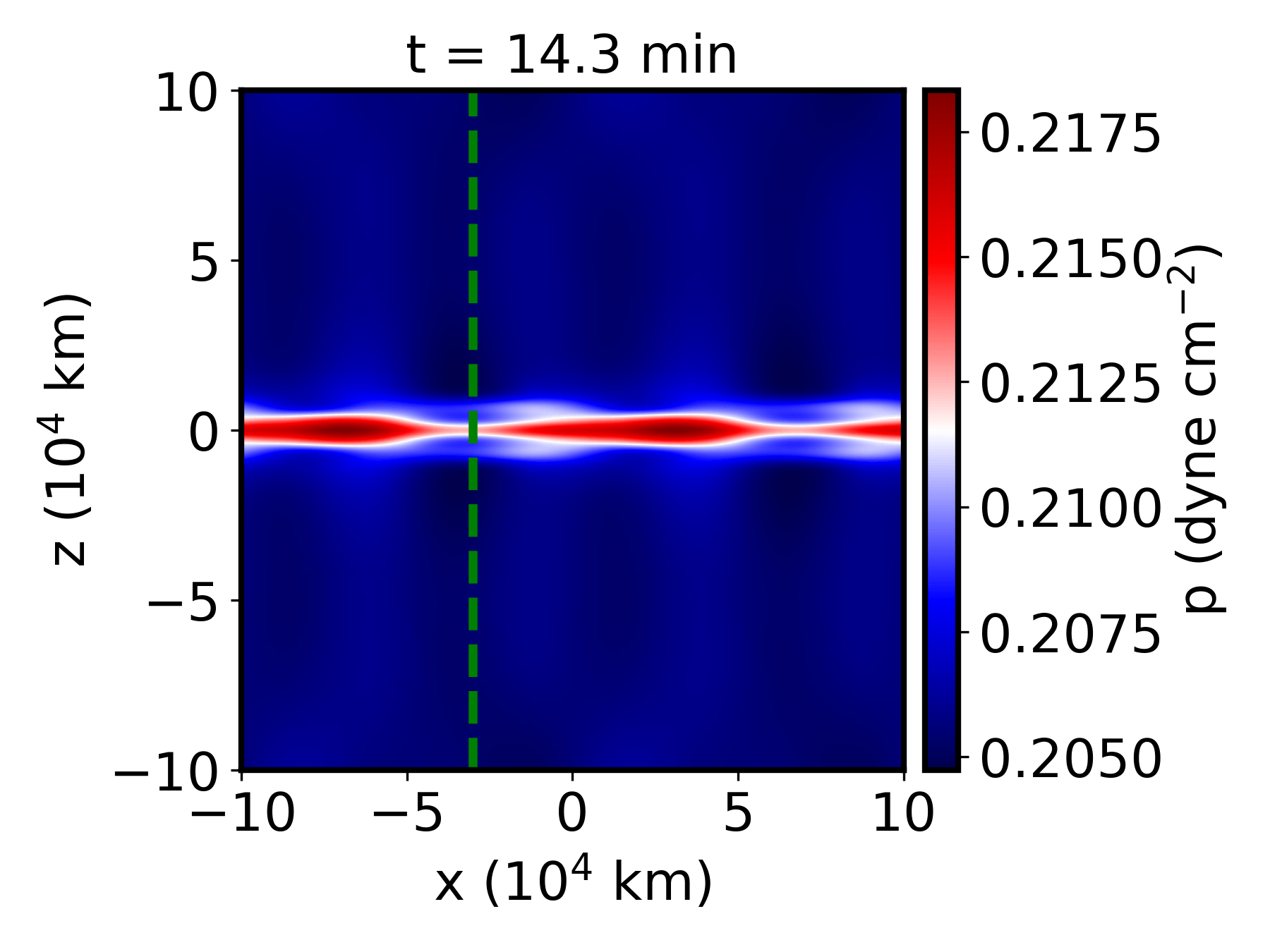}
    \caption{at $y=0$}
    \label{fig:p_xz_t010}
\end{subfigure}
\hspace{0.5 cm}
\begin{subfigure}{0.24\textwidth}
    \includegraphics[width=1.1\textwidth]{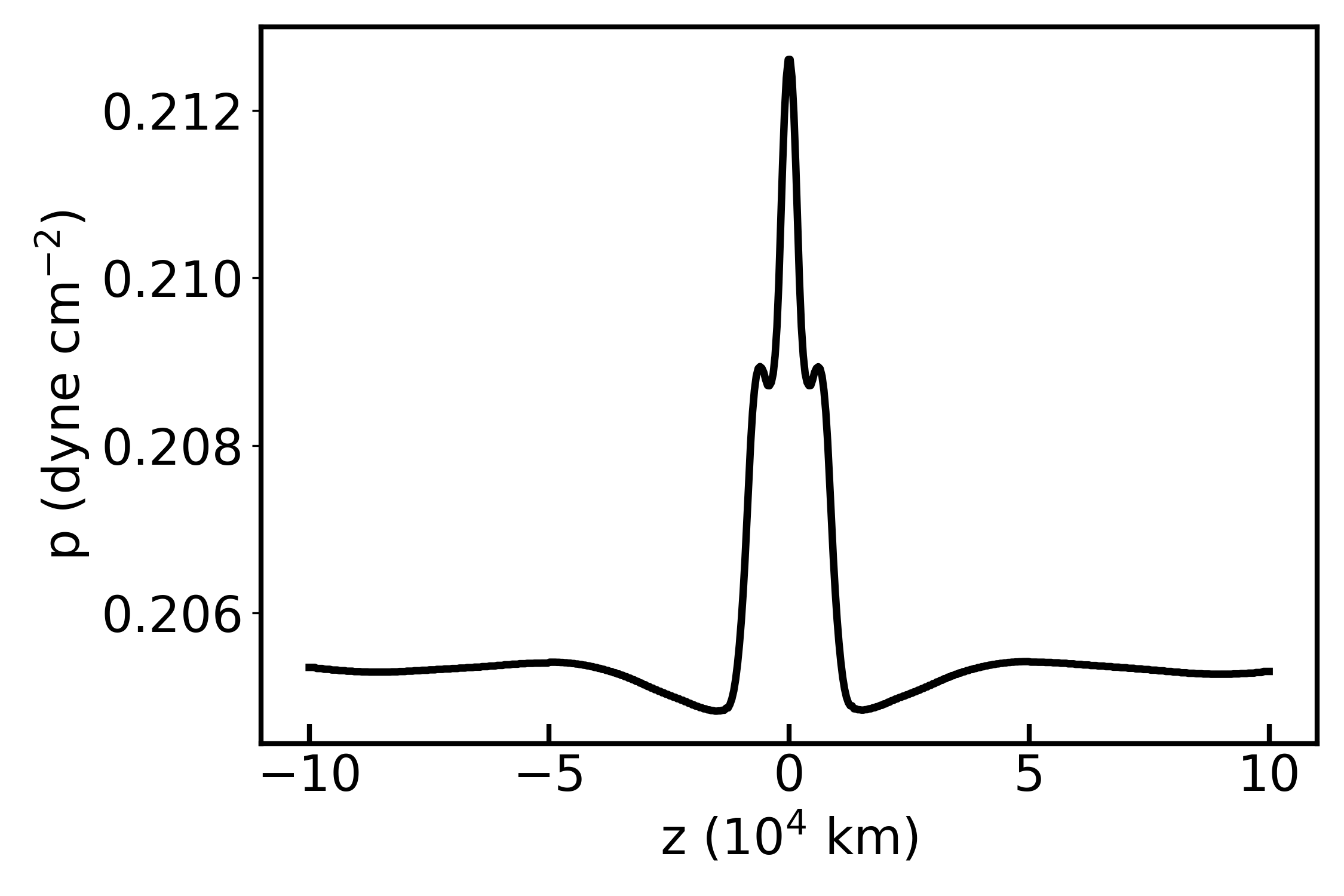} 
    \caption{$t=14.3$ min}\label{fig:p_zcut_t010}
\end{subfigure}
\caption{The growth of the tearing mode around the current sheet plane. We show data at $t=14.3$ min, where (a) represents the evolution of the $B_z$ component (perpendicular to the current sheet, $B_z$ was initially zero, with no magnetic field perturbation along the $z$ direction), (b) plasma pressure in the $x-z$ plane (at $y=0$). (c) represents the variation of the plasma pressure along the green dashed vertical line in figure (b).}
\label{fig:tearing_analysis}
\end{figure*}

\begin{figure*}[hbt!]
\centering
\begin{subfigure}{0.38\textwidth}
    \includegraphics[width=1.1\textwidth]{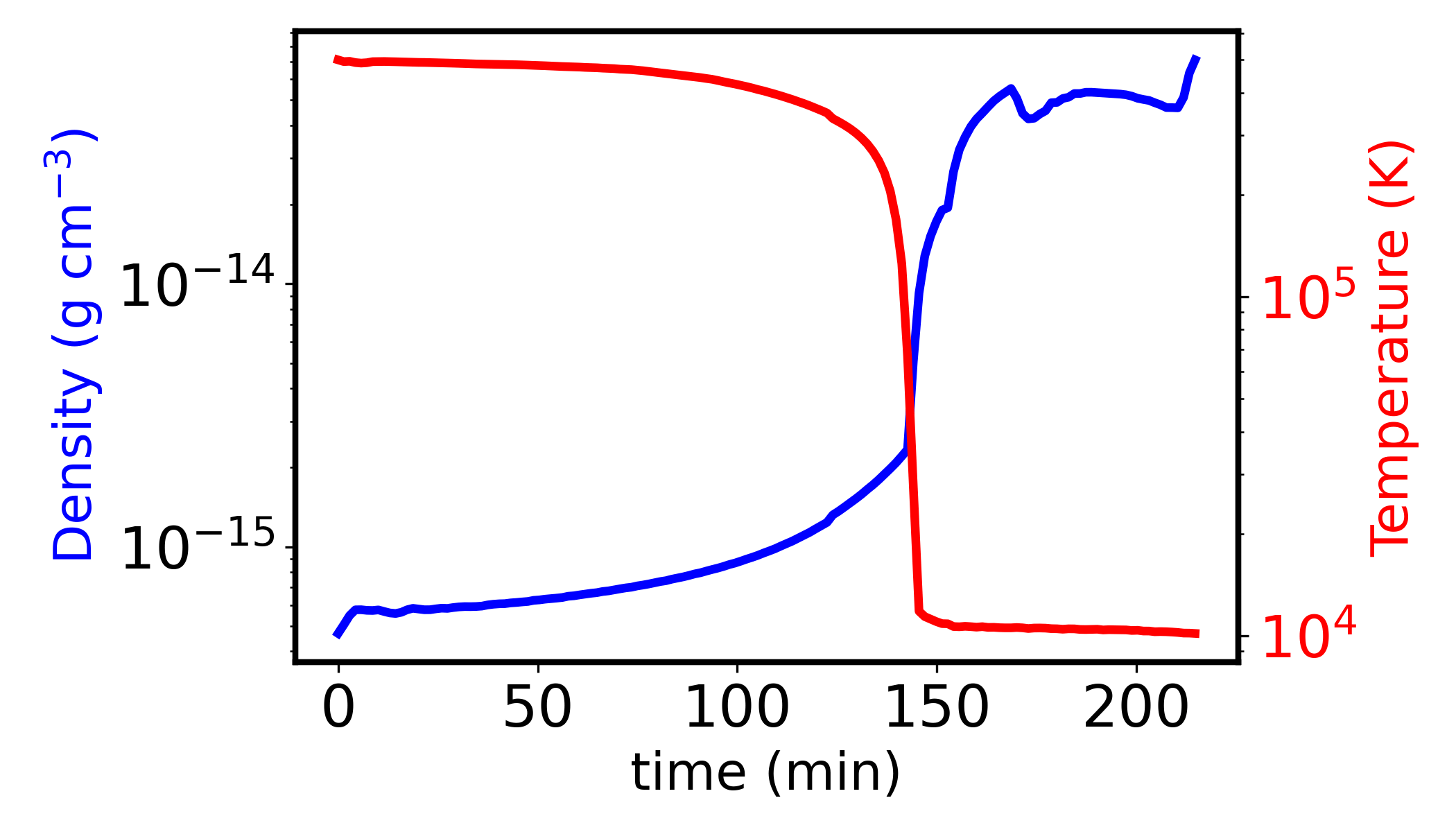}
    \caption{}
    \label{fig:den_tem_time}
\end{subfigure}
\hspace{1.2 cm}
\begin{subfigure}{0.3\textwidth}
    \includegraphics[width=1.1\textwidth]{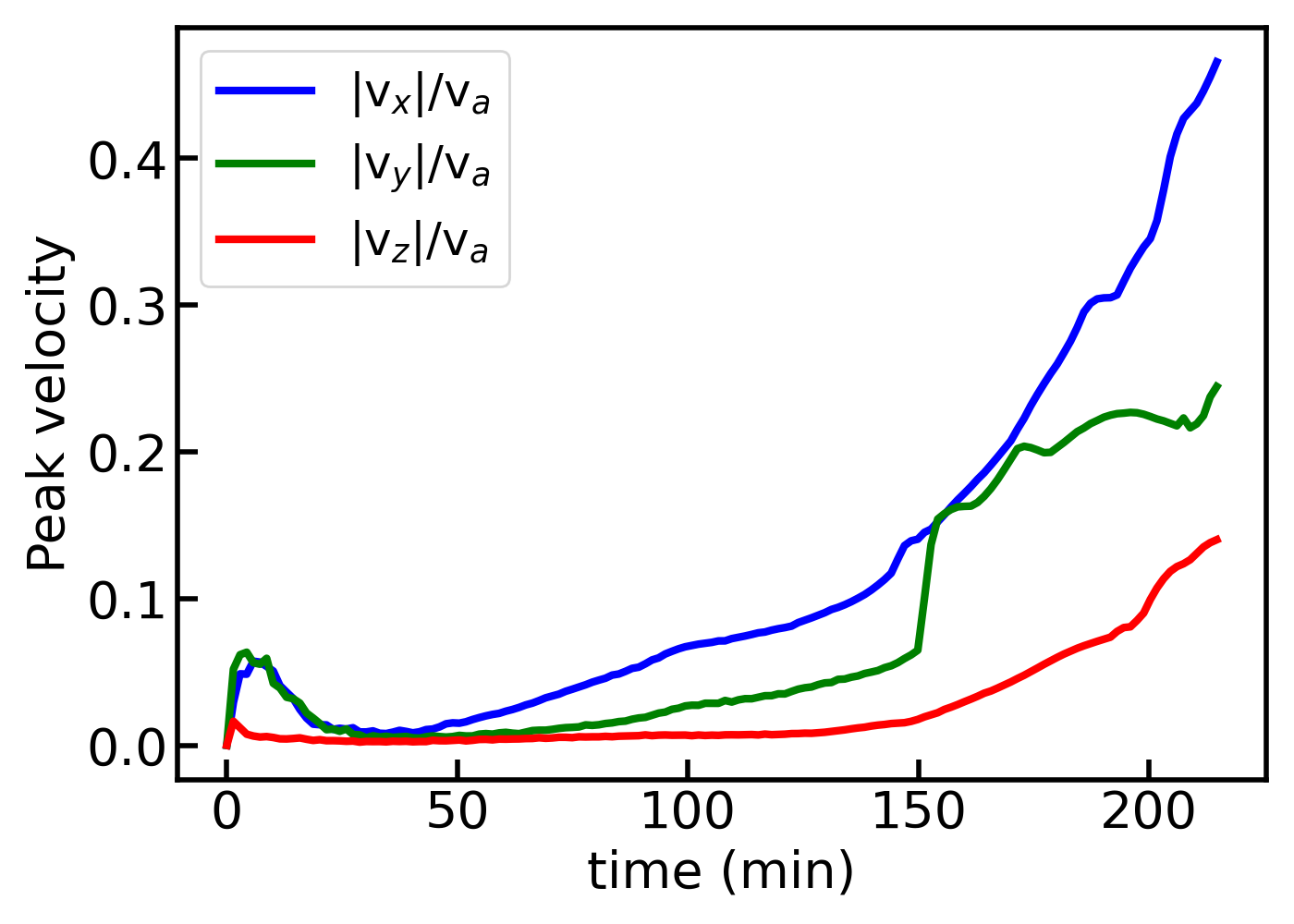} 
    \caption{} \label{fig:peak_vel}
\end{subfigure}
    \caption{Temporal variation of the (a) instantaneous maximal plasma density (blue line), and minimum temperatures (red line), and (b) instantaneous absolute peak velocities. The sharp rise of the density, and drop of minimum temperature over two orders of magnitude at $\approx$ 150 min signals the runaway thermal instability causing local condensations, when the absolute peak velocities, $v_x$, $v_y$, and $v_z$ also rise sharply. Here, the velocities are scaled in terms of Alfv{\'e}n velocity, $v_a \approx 261$ km s$^{-1}$.}
    \label{fig:dens-temp-vel-TS}
\end{figure*}

\begin{figure}[hbt!]
    \centering
    \includegraphics[width=0.4 \textwidth]{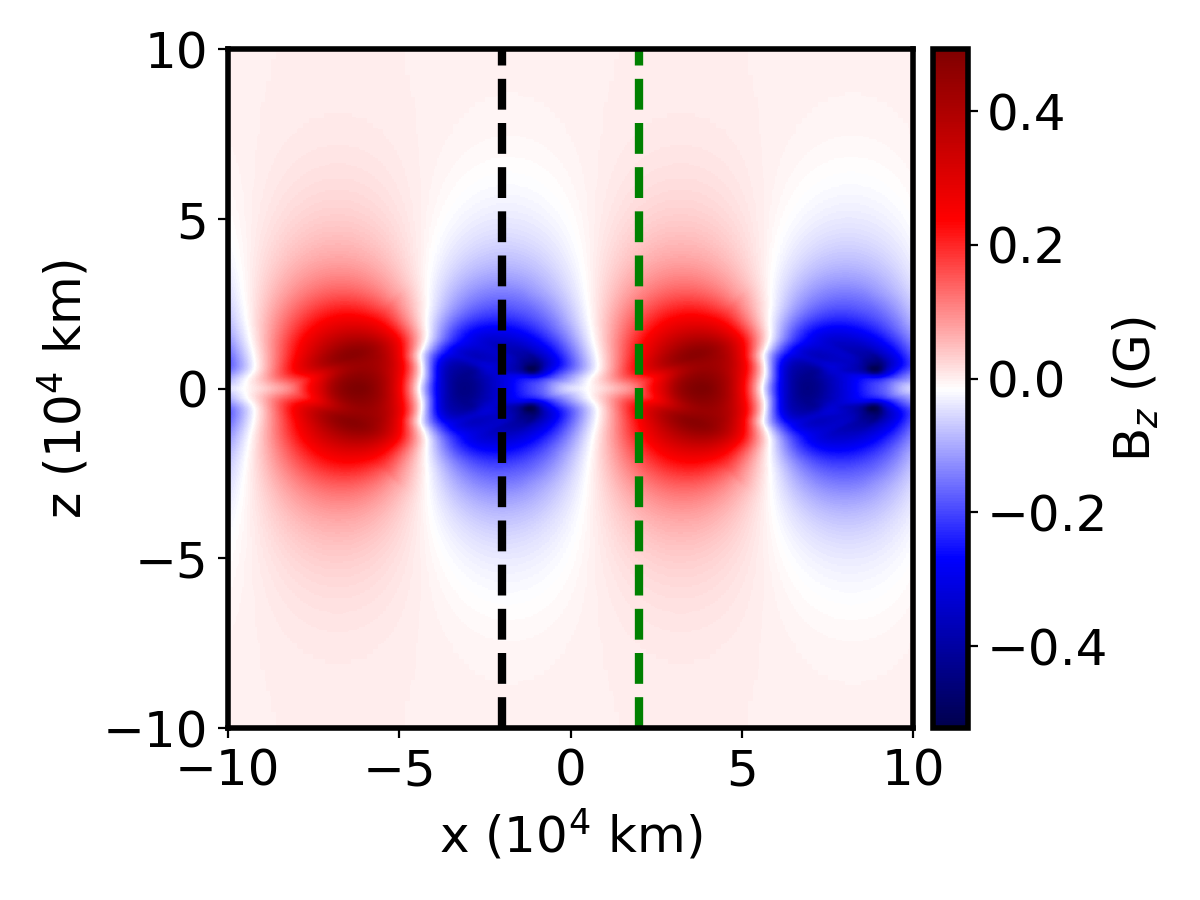}
    \centering
    \includegraphics[width=0.35 \textwidth]{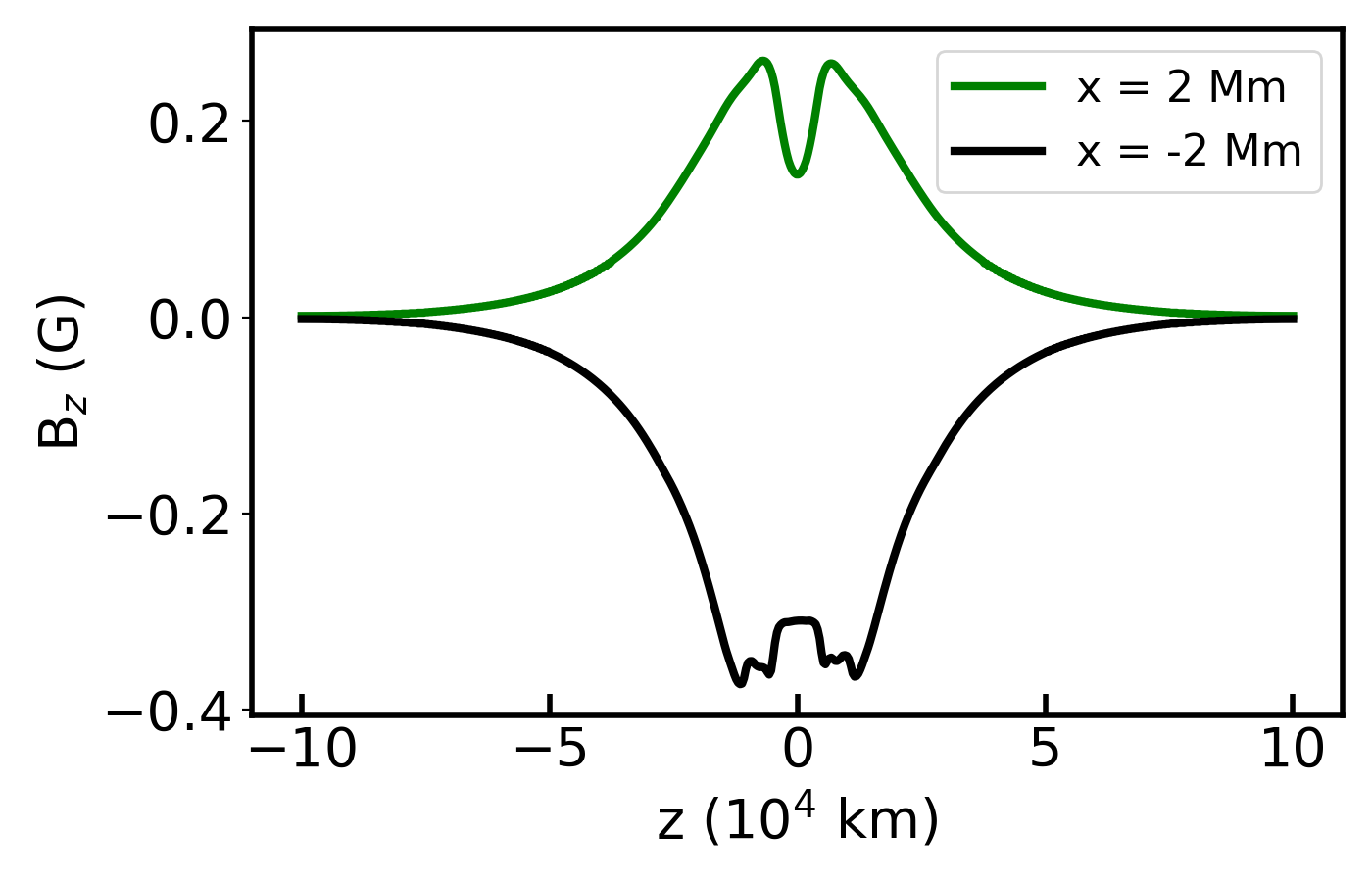}
    \caption{Signature of tearing modes around the fragmented current sheet plane at $t=207.5$ min. Top panel represents the variation of $B_z$ along $x-z$ at $y=0$ plane. Bottom panel shows variation of $B_z$ along the dashed vertical lines in the top panel at $x=\pm 2$ Mm.}
    \label{fig:tearing_later}
\end{figure}

\subsection{Global evolution}\label{global evolution}

The spatial distribution of current density squared $J^2$, as well as representative corresponding field line evolutions are shown in Fig.~\ref{fig:jsqr}. The equilibrium configuration of the current sheet (at $t=0$) is formed due to the fact that the magnetic field shears across $z=0$ as given by Eqs. \ref{Bx}, \ref{By}, and \ref{Bz}. The initial current distribution is mostly oriented in the $x-y$ plane and concentrated around $z=0$, with its main contribution from $J_x=-dB_y/dz$ and $J_y=dB_x/dz$ (while $J_z$ is purely from the perturbed field). Due to the magnetic field perturbations, given by Eqs. \ref{dbx} and \ref{dby}, which are (mainly) confined near the $z=0$ plane and extend along the $x-y$ directions, linear resistive tearing modes start to develop leading to the disintegration of the current sheet. The inhomogeneities of $J^2$ in the current sheet plane due to the multimode perturbation ($n_1=4$, $n_2=2$) appear at the initial stage of the evolution as shown in Fig. \ref{fig:jsqr_t010}. Magnetic reconnections lead to pronounced magnetic topology changes, modifying the current sheet as shown in Fig.~\ref{fig:jsqr_t145}. We see the perturbed field lines near $z=0$ (see Fig.~\ref{fig:fl_t003}) turn into extended flux rope-like structures while the system evolves (see Fig. \ref{fig:fl_t145}), yet the planar ($x$-oriented) field lines away from the current sheet plane remain unperturbed. The signature of flux ropes in the current density distribution in Fig.~\ref{fig:jsqr_t145} can be clearly noticed. In a 2D setup, these would be the familiar magnetic islands or plasmoids, due to development of tearing modes in the current sheet. We notice the self consistent development of $B_z$ due to the perturbed fields around the current sheet plane as shown in Fig. \ref{fig:bz_xz_t010} (note that the initial $B_z$ was set to zero with no magnetic field perturbation along $z$). Fig.~\ref{fig:p_xz_t010} represents the plasma pressure distribution at the $x-z$ plane (at $y=0$), while the variation along the vertical green dashed line is shown in Fig.~ \ref{fig:p_zcut_t010}, which shows the expected linear eigenmode structure of the tearing mode, seen as the kinks in the pressure distribution around $z=0$. This implies the development of the tearing modes at the initial stage of the current sheet evolution. The nonlinearly developing tearing evolution creates density perturbations in the surroundings of the current sheet, and the radiative cooling (in combination with thermal conduction)  becomes dominant over the constant background heating in localized regions of the domain. This triggers the cooling of those regions, which in turn condenses the regions even more. Hence, a runaway process starts where tearing and thermal modes reinforce each other, which causes the spontaneous growth of density and temperature inhomogeneities in the surroundings of the current sheet. Note that our box is periodic along $x$ and $y$, so field aligned thermal conduction does not play a big role in the heating-cooling misbalance of the system (in the sense that it can not lead to heat fluxes down into lower-lying chromospheric regions: we have no stratification here). It just tries to homogenize the temperature along field lines, in competition with local resistive heating. 

\begin{figure*}[hbt!]
\centering
\begin{subfigure}{0.28\textwidth}
    \includegraphics[width=1.1\textwidth]{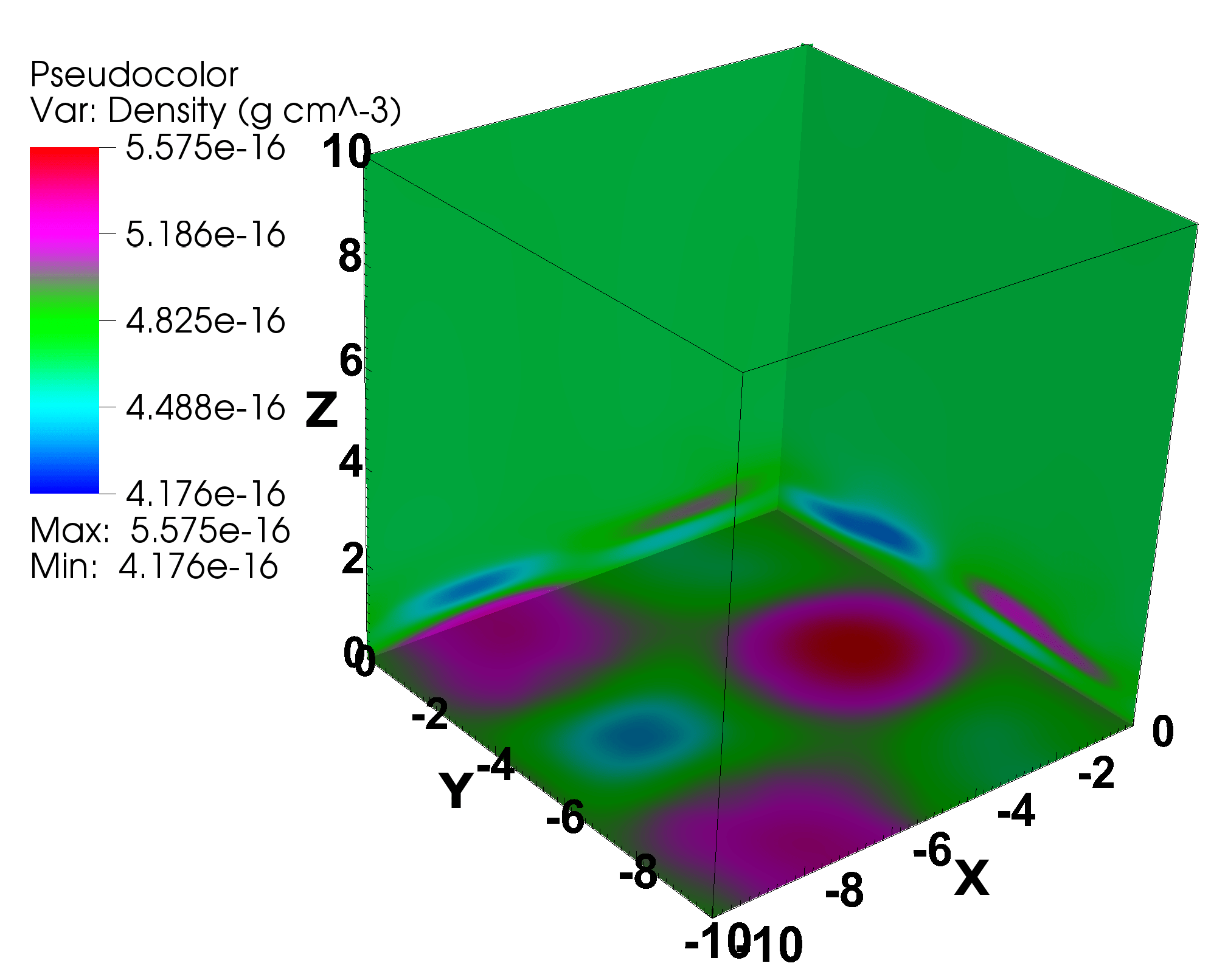}
    \caption{t=14.3 min}
    \label{fig:denslice_t010}
\end{subfigure}
\hspace{0.5 cm}
\begin{subfigure}{0.28\textwidth}
    \includegraphics[width=1.1\textwidth]{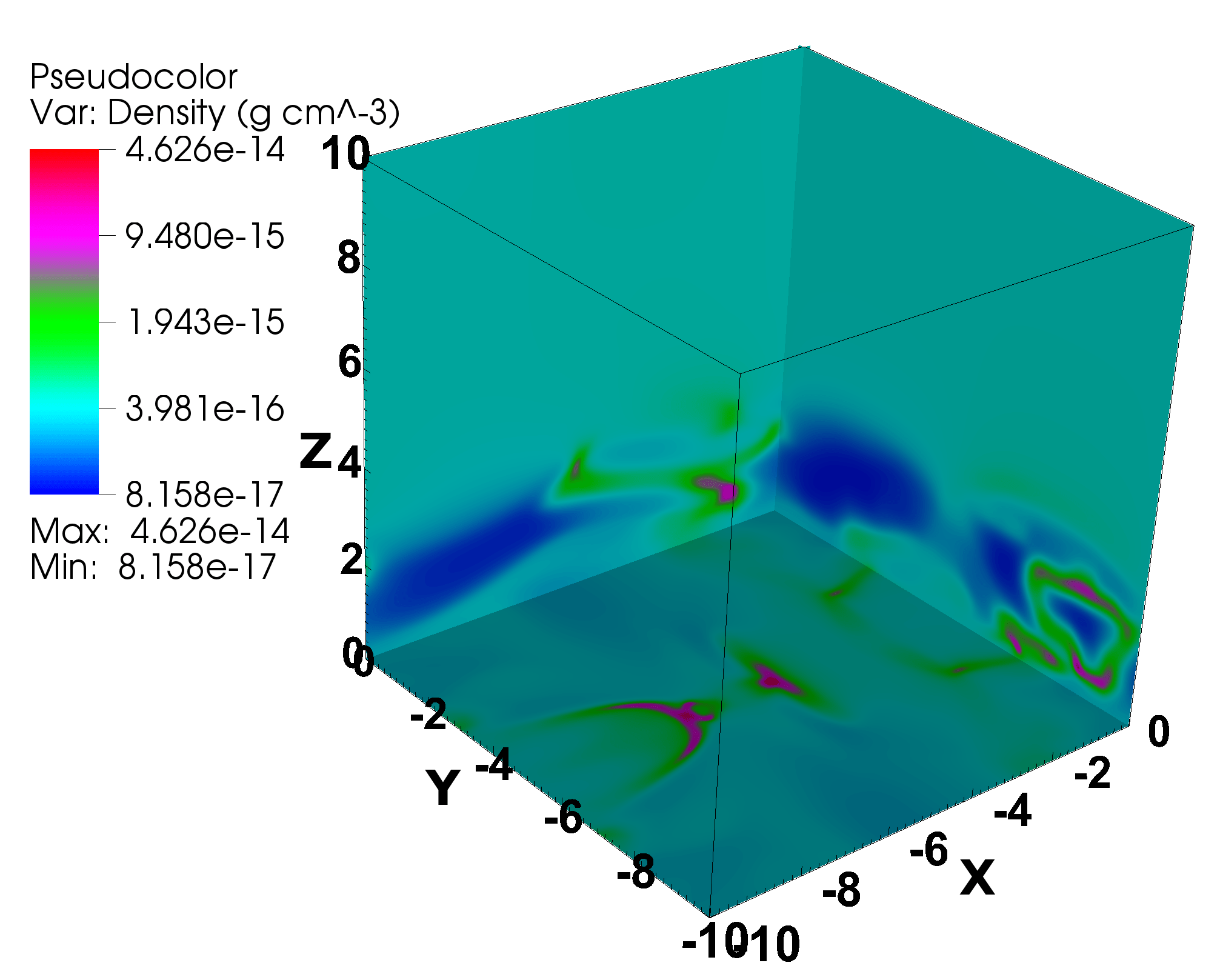}
    \caption{t=207.5 min}
    \label{fig:denslice_t145}
\end{subfigure}
\hspace{0.5 cm}
\begin{subfigure}{0.32\textwidth}
    \includegraphics[width=1.1\textwidth]{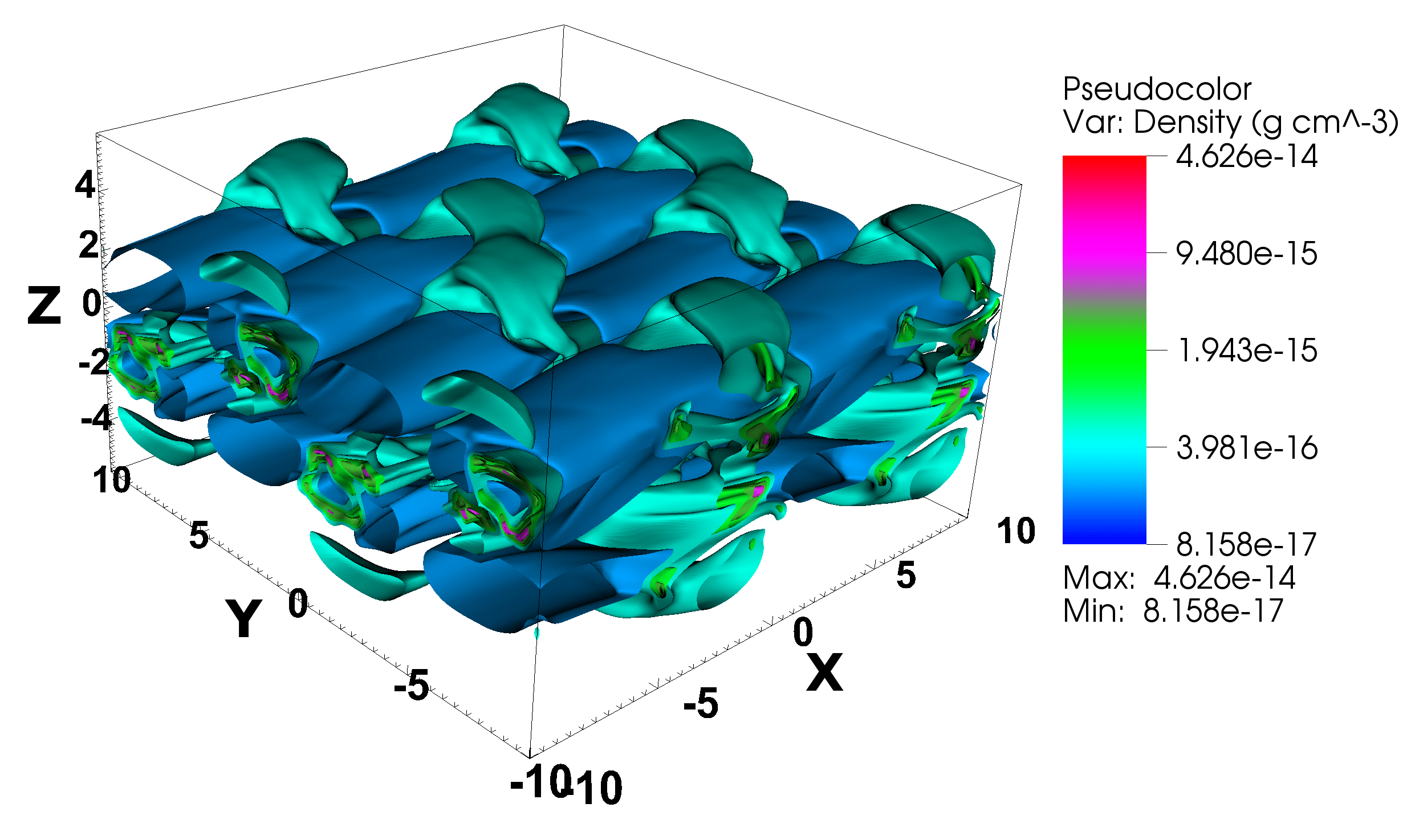}
    \caption{t=207.5 min}
    \label{fig:den3d_t145}
\end{subfigure}
\newline
\centering
\begin{subfigure}{0.28\textwidth}
    \includegraphics[width=1.1\textwidth]{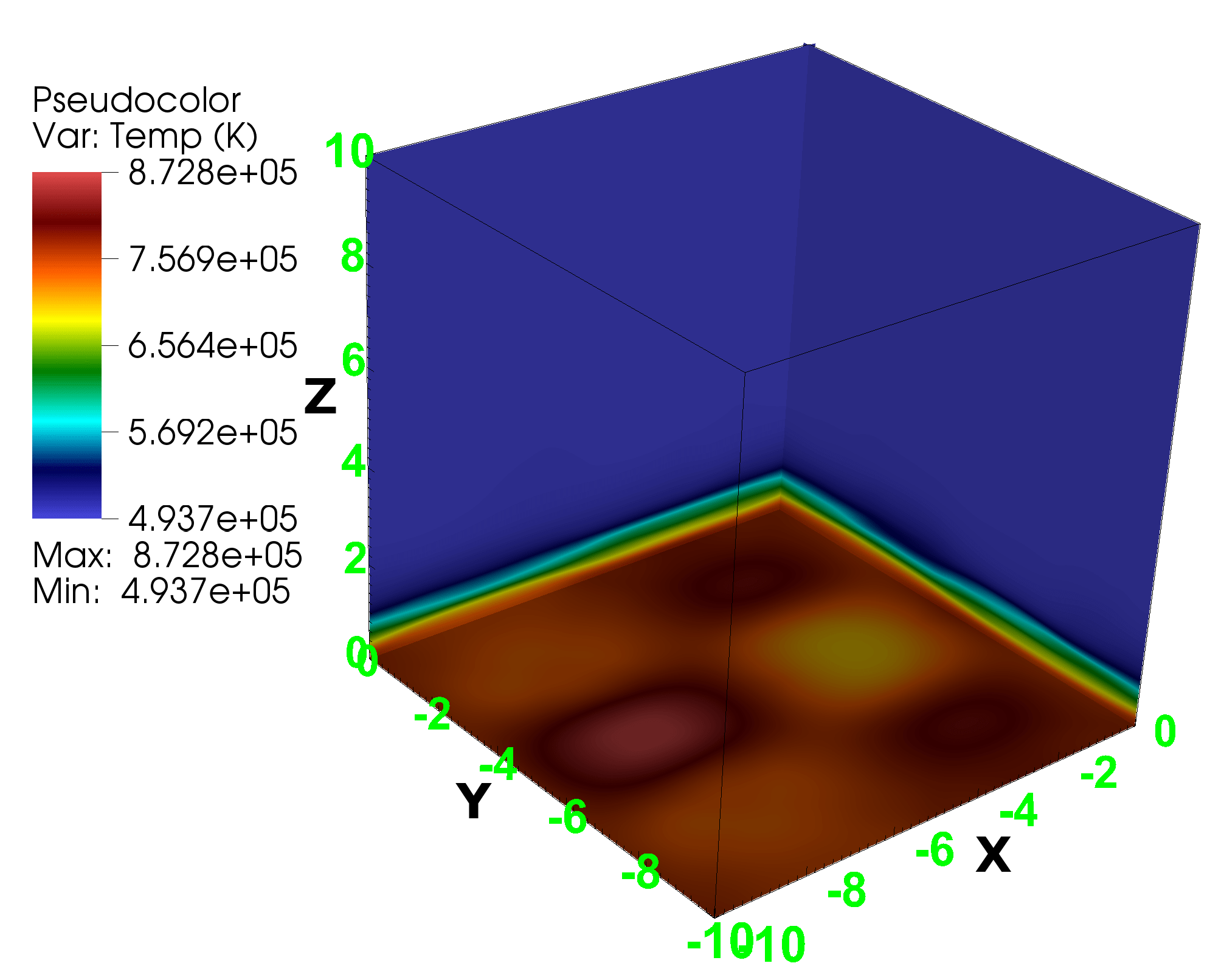}
    \caption{t=14.3 min}
    \label{fig:templice_t010}
\end{subfigure}
\hspace{0.5 cm}
\begin{subfigure}{0.28\textwidth}
    \includegraphics[width=1.1\textwidth]{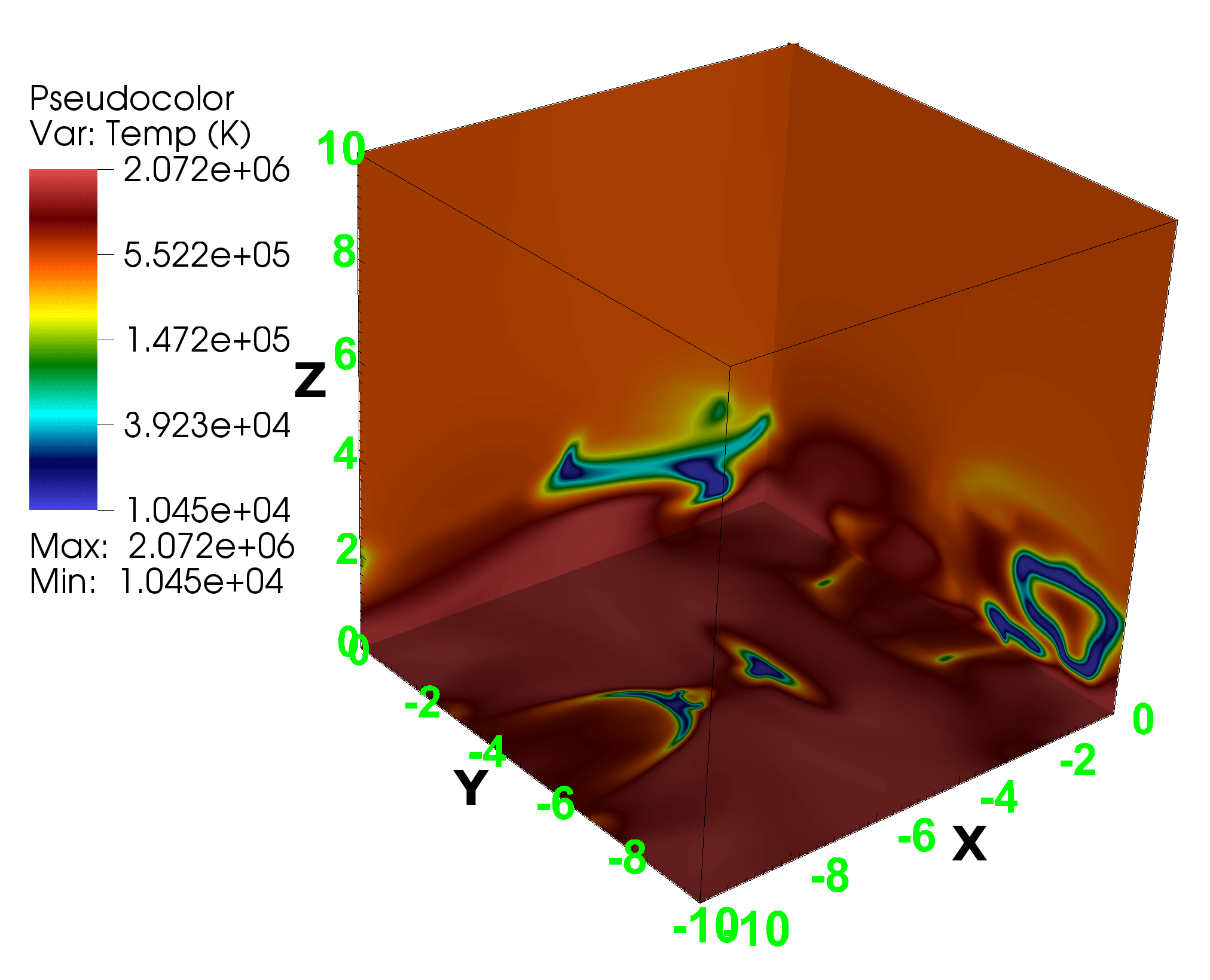}
    \caption{t=207.5 min}
    \label{fig:templice_t145}
\end{subfigure}
\hspace{0.5 cm}
\begin{subfigure}{0.32\textwidth}
    \includegraphics[width=1.1\textwidth]{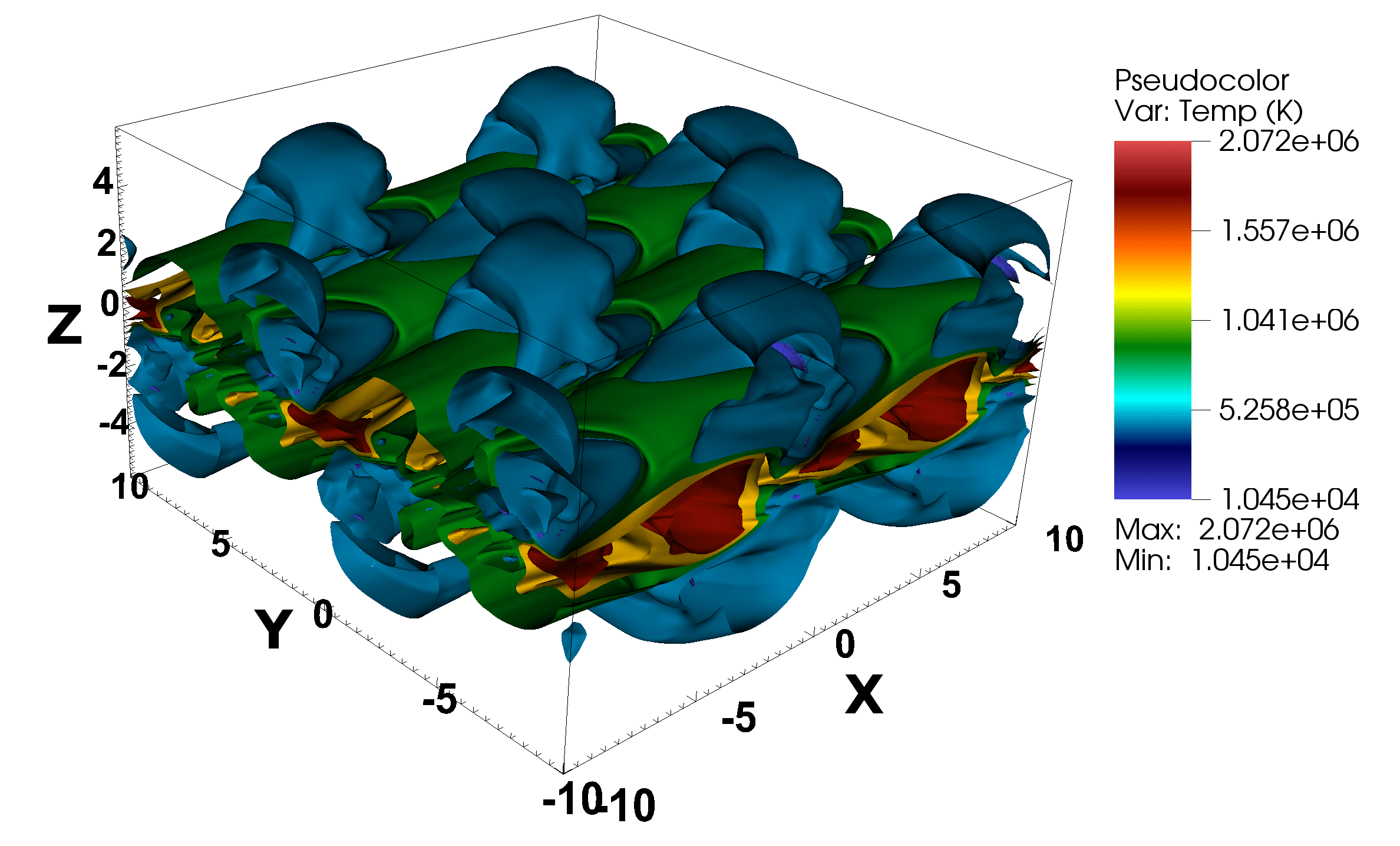}
    \caption{t=207.5 min}
    \label{fig:tem3d_t145}
\end{subfigure}
\caption{Spatial distribution of density (top panel) and temperature (bottom) in the 3D domain, where the distances along $x, y,$ and $z$ directions are in units of $10^4$ km. Density (a and b), and temperature (d and e) distribution along three orthogonal slices along $x=0$, $y=0$, and $z=0$ planes for two different times, $t=14.3$ and $207.5$ min are shown. (c) and (f) represent isosurface views (five isosurfaces ranging from minimum to maximum values) on density and temperature, respectively. Panels (a) and (d) are the early phase of the evolution, where the density and temperature inhomogeneities appear around the current sheet plane due to the multi mode magnetic field perturbation. Panel (b) and (c) illustrate where high density structures appear, cospatial with cool ($\sim 10^4$ K) regions in (e) and (f). (An animation is available online.)}
\label{fig:den-temp-3dview}
\end{figure*}

\begin{figure*}[hbt!]
\centering
\begin{subfigure}{0.38\textwidth}
    \includegraphics[width=1.1\textwidth]{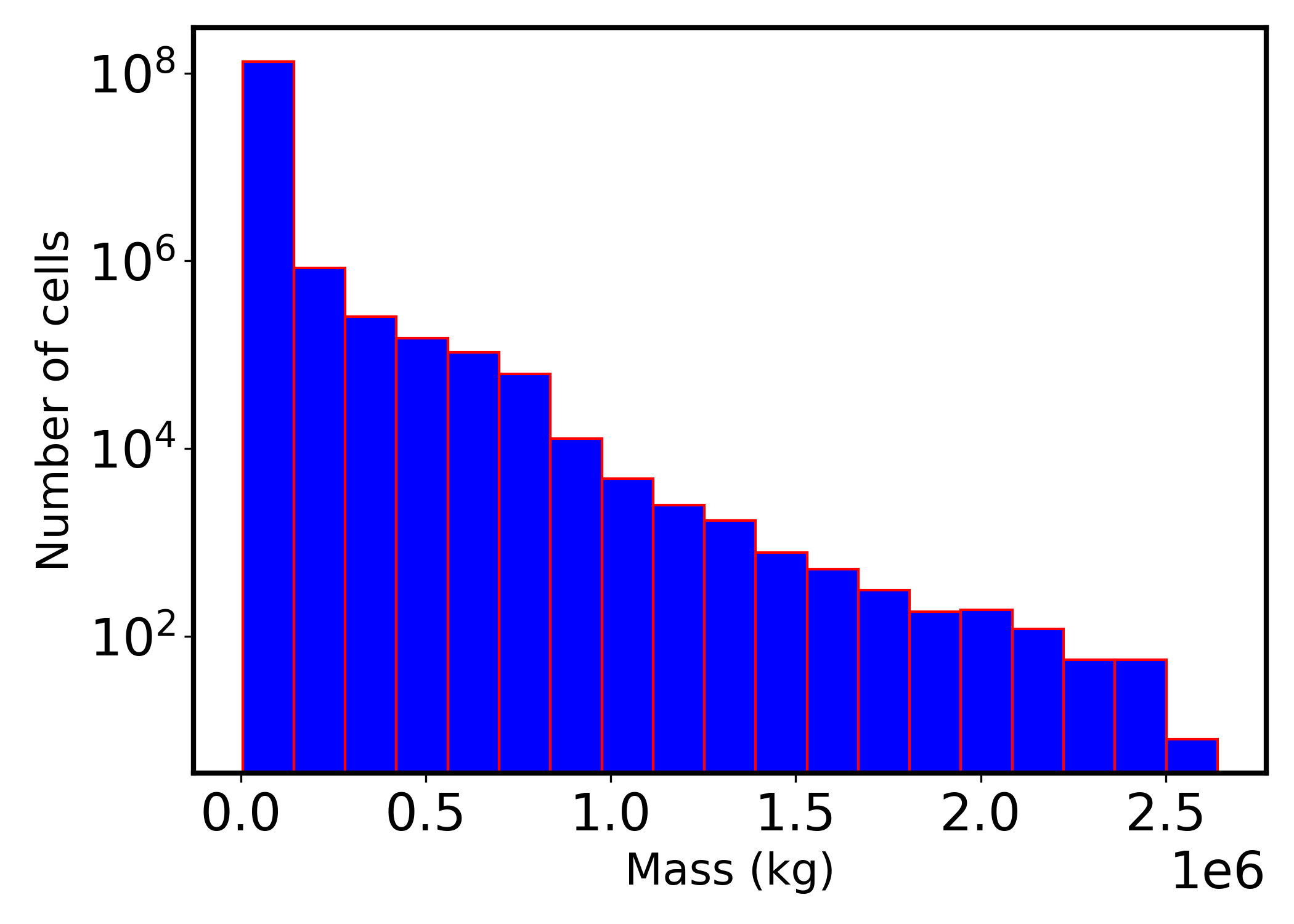}
    \caption{}
    \label{fig:mass_dist_t145}
\end{subfigure}
\hspace{0.8 cm}
\begin{subfigure}{0.38\textwidth}
    \includegraphics[width=1.1\textwidth]{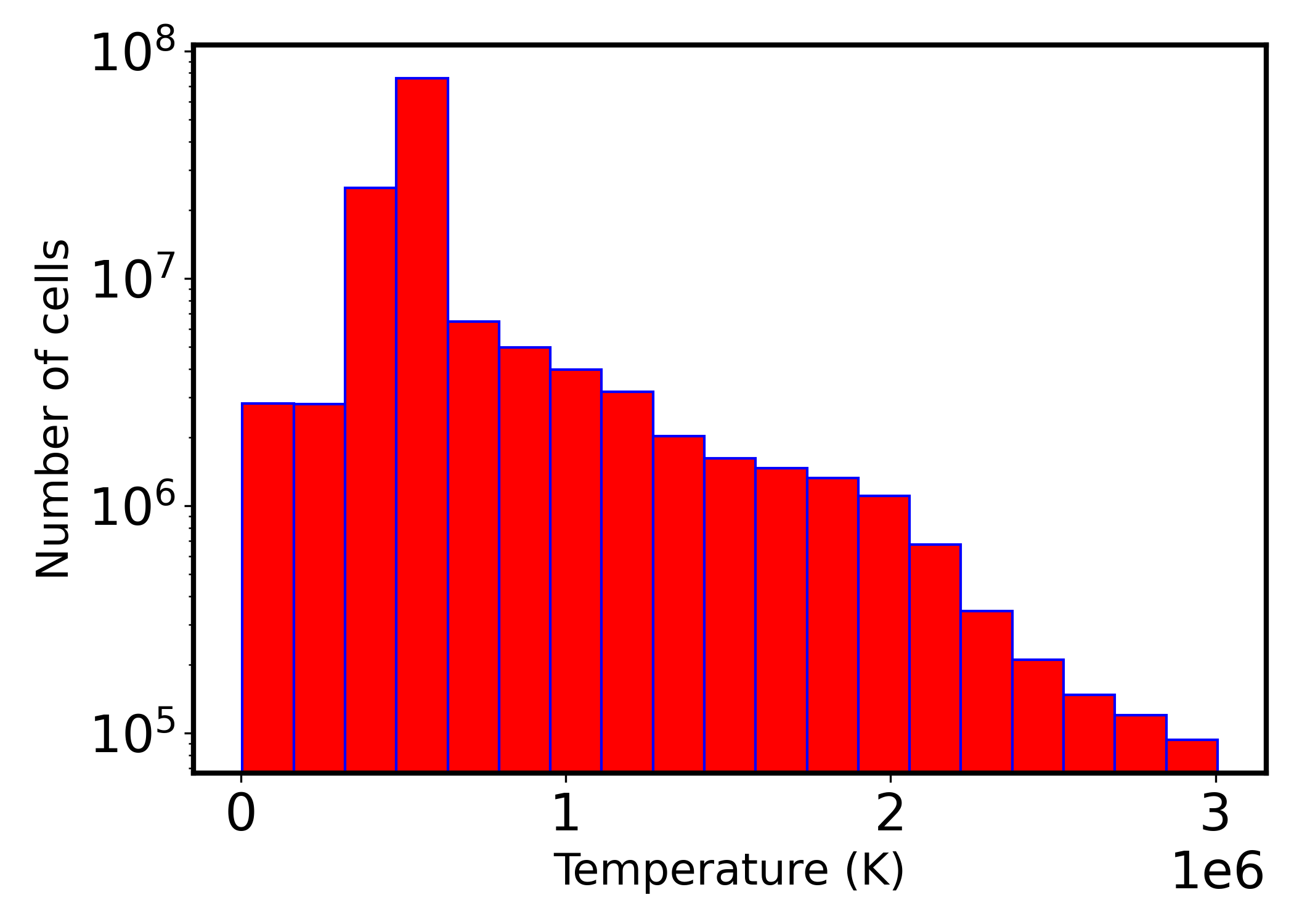} 
    \caption{} \label{fig:temp_dist_t145}
\end{subfigure}
    \caption{Histograms for (a) mass, and (b) temperature distribution for $t=207.5$ min. The cells containing the minimum and maximum masses are $4.8 \times 10^4$ and $2.7 \times 10^6$ kg respectively, and the temperature minima and maxima for the cells are 5000 K and 3.1 MK repectively. The number of bins for (a) and (b) are 20 with bin sizes $1.38 \times 10^5$ kg and 0.15 MK respectively.}
    \label{fig:mass-temp-dist}
\end{figure*}

\begin{figure*}[hbt!]
\centering
\begin{subfigure}{0.3\textwidth}
    \includegraphics[width=1.1\textwidth]{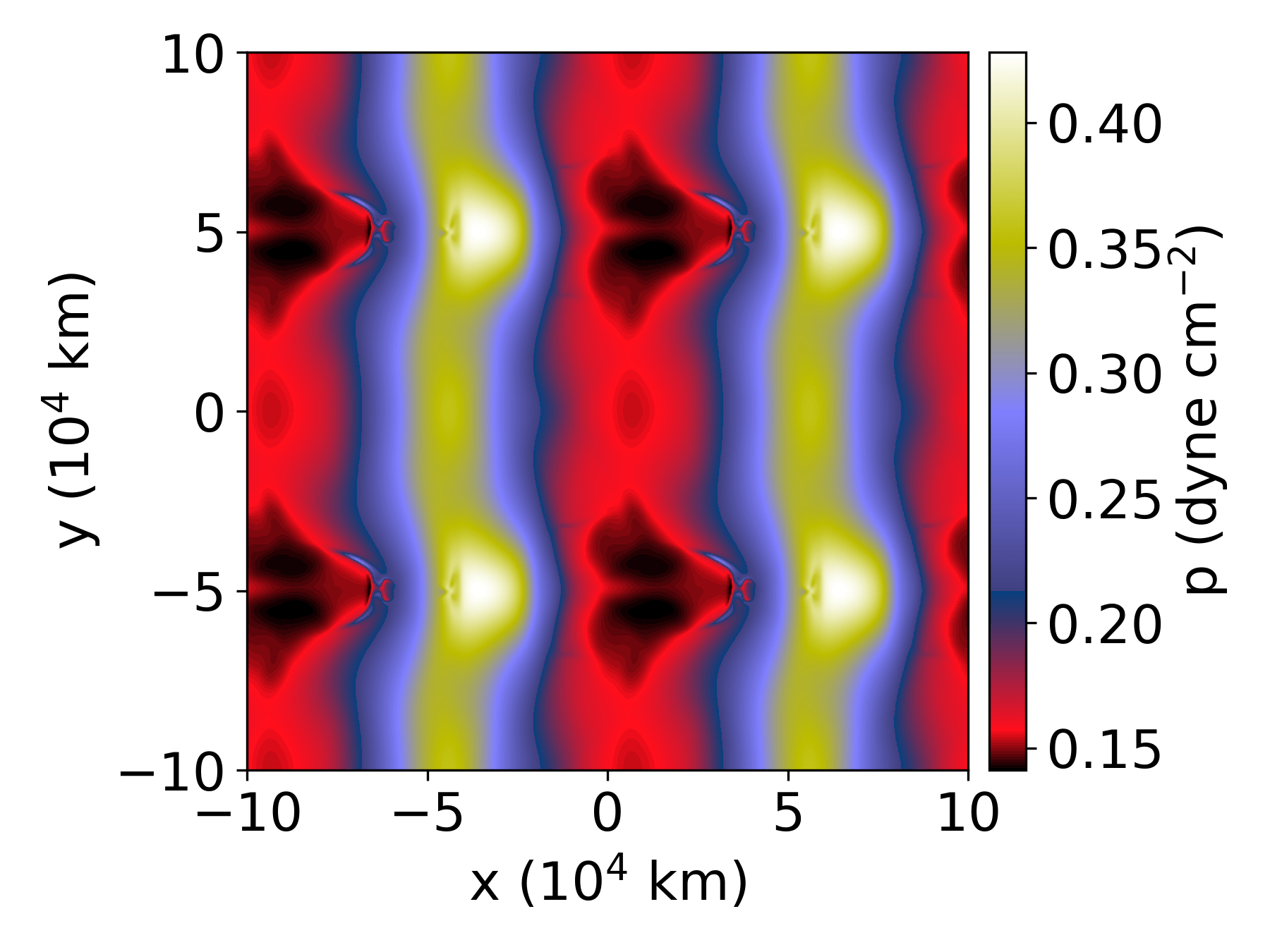}
    \caption{at $z=0$}
    \label{fig:p_xy_t145}
\end{subfigure}
\hspace{0.5 cm}
\begin{subfigure}{0.3\textwidth}
    \includegraphics[width=1.1\textwidth]{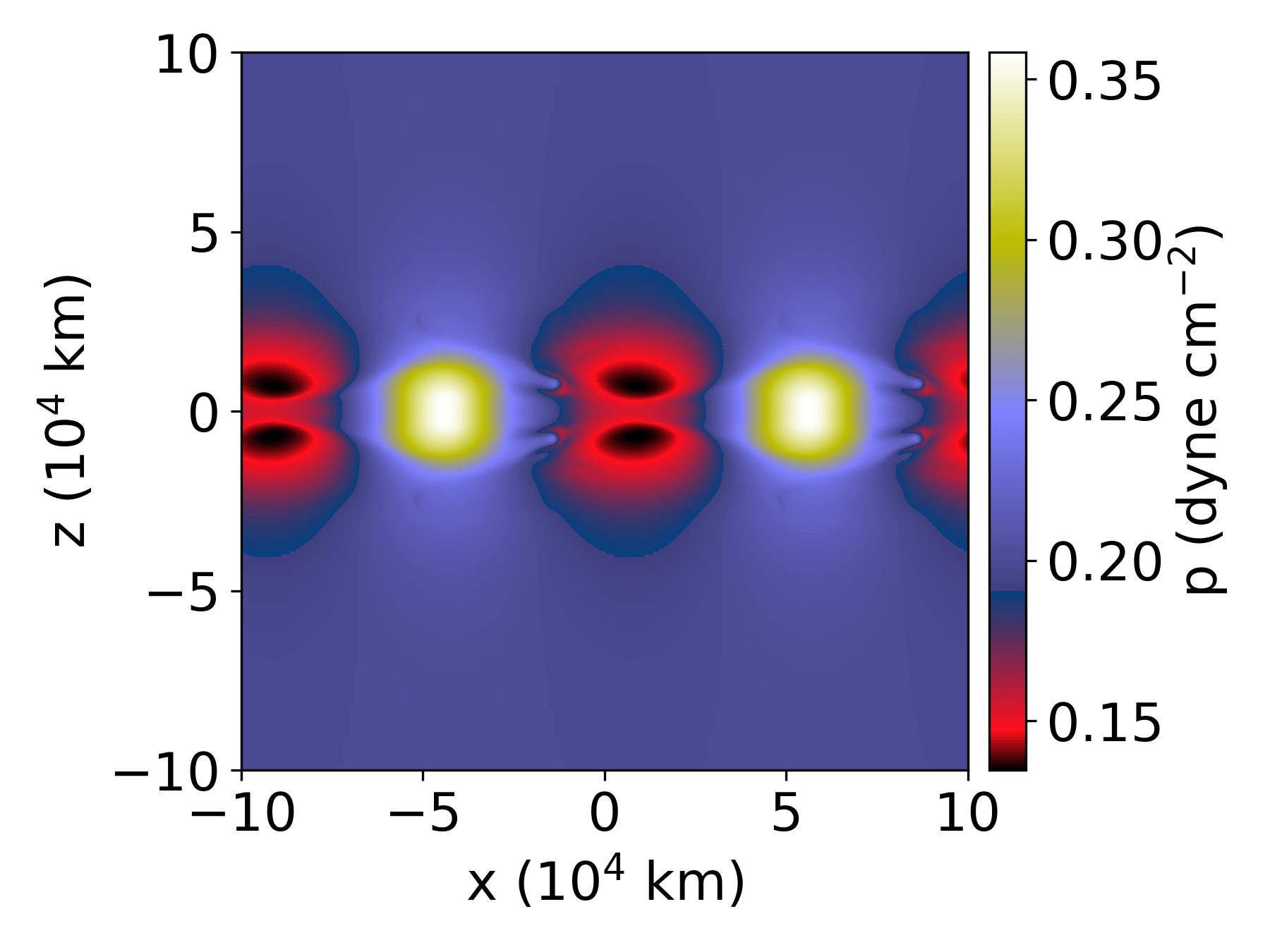}
    \caption{at $y=0$}
    \label{fig:p_xz_t145}
\end{subfigure}
\hspace{0.5 cm}
\begin{subfigure}{0.3\textwidth}
    \includegraphics[width=1.1\textwidth]{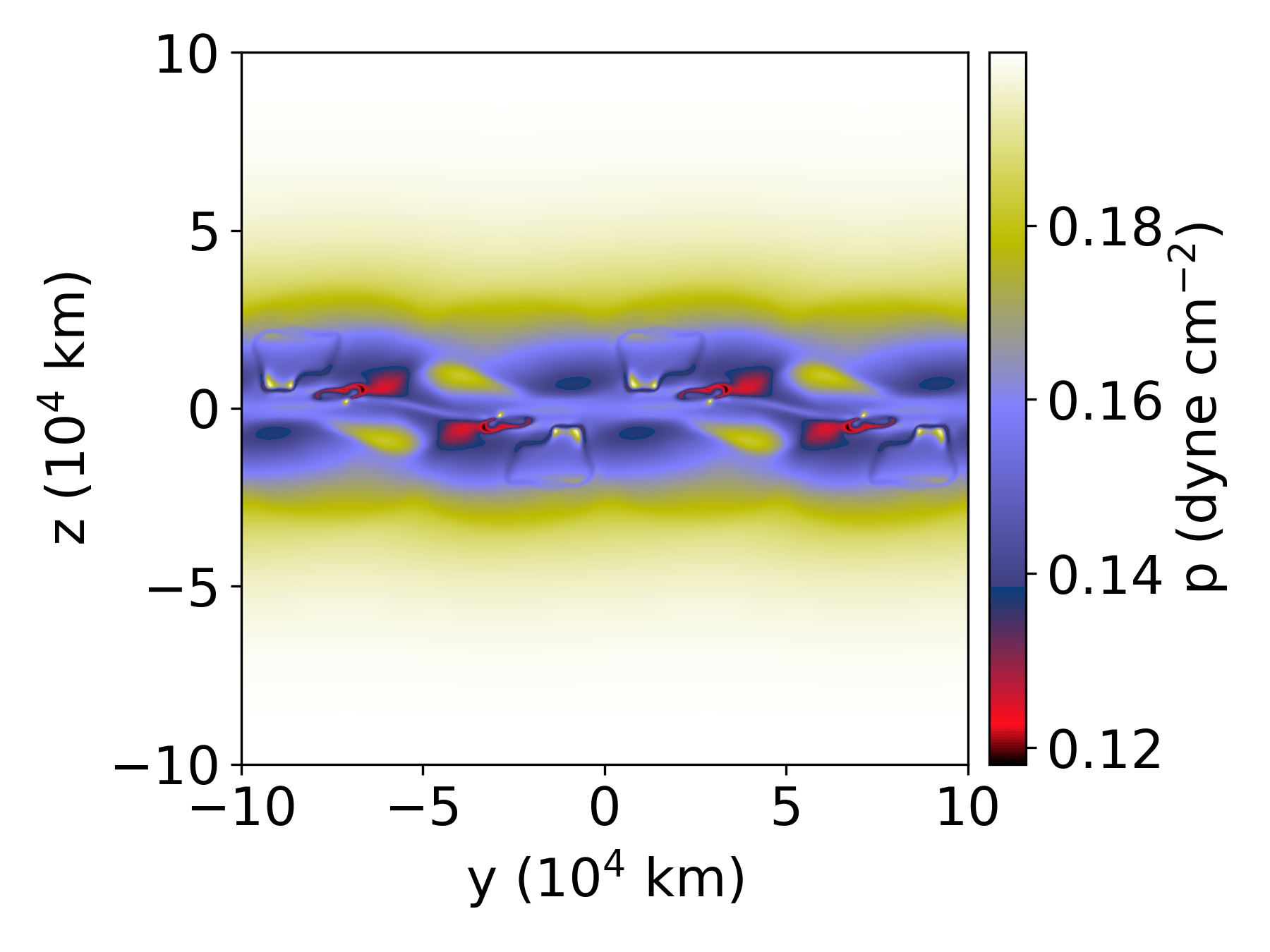}
    \caption{at $x=0$}
    \label{fig:p_yz_t145}
\end{subfigure}
\newline
\centering
\begin{subfigure}{0.3\textwidth}
    \includegraphics[width=1.1\textwidth]{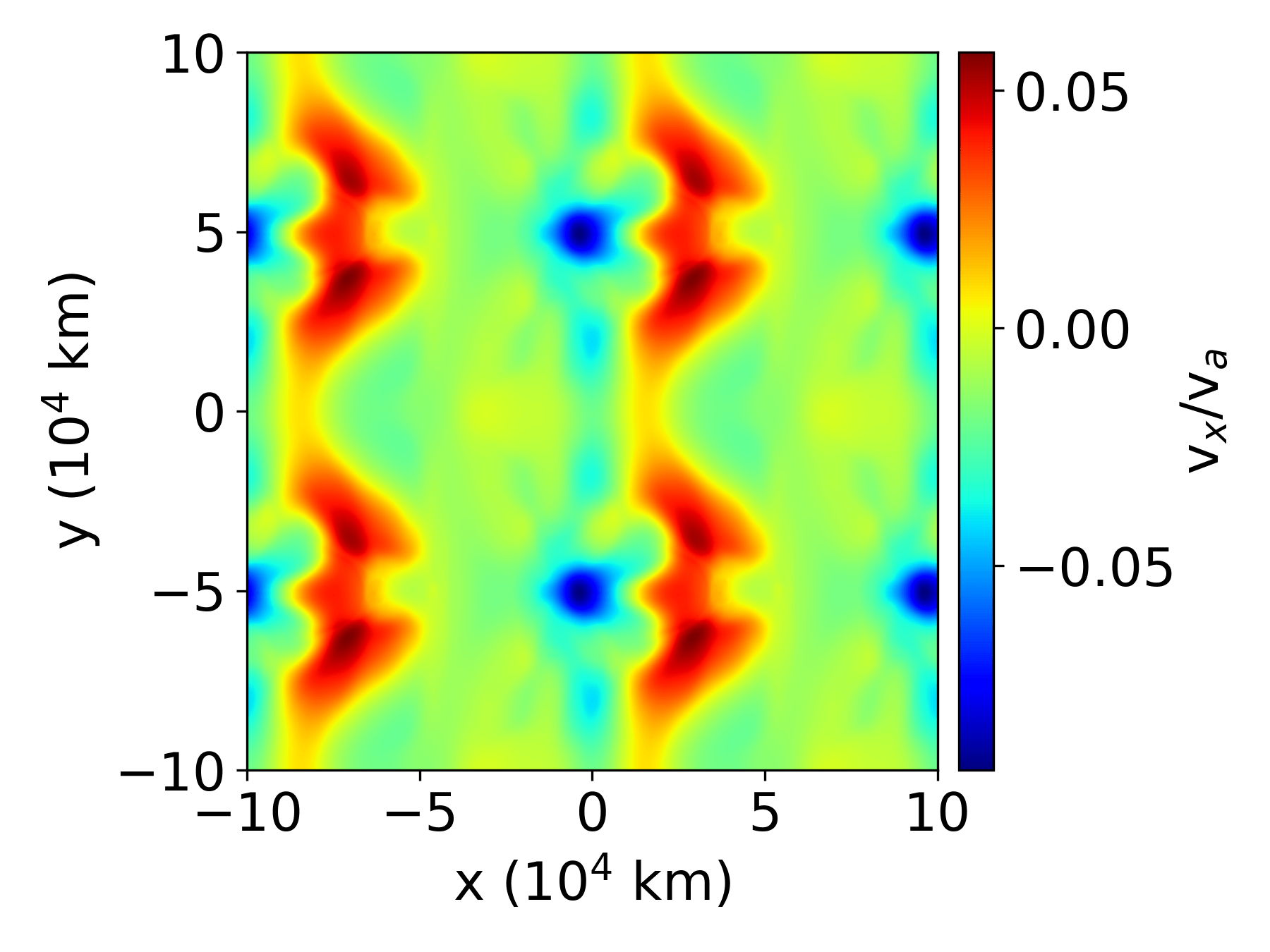}
    \caption{at $z=0$}
    \label{fig:vx_xy_t145}
\end{subfigure}
\hspace{0.5 cm}
\begin{subfigure}{0.3\textwidth}
    \includegraphics[width=1.1\textwidth]{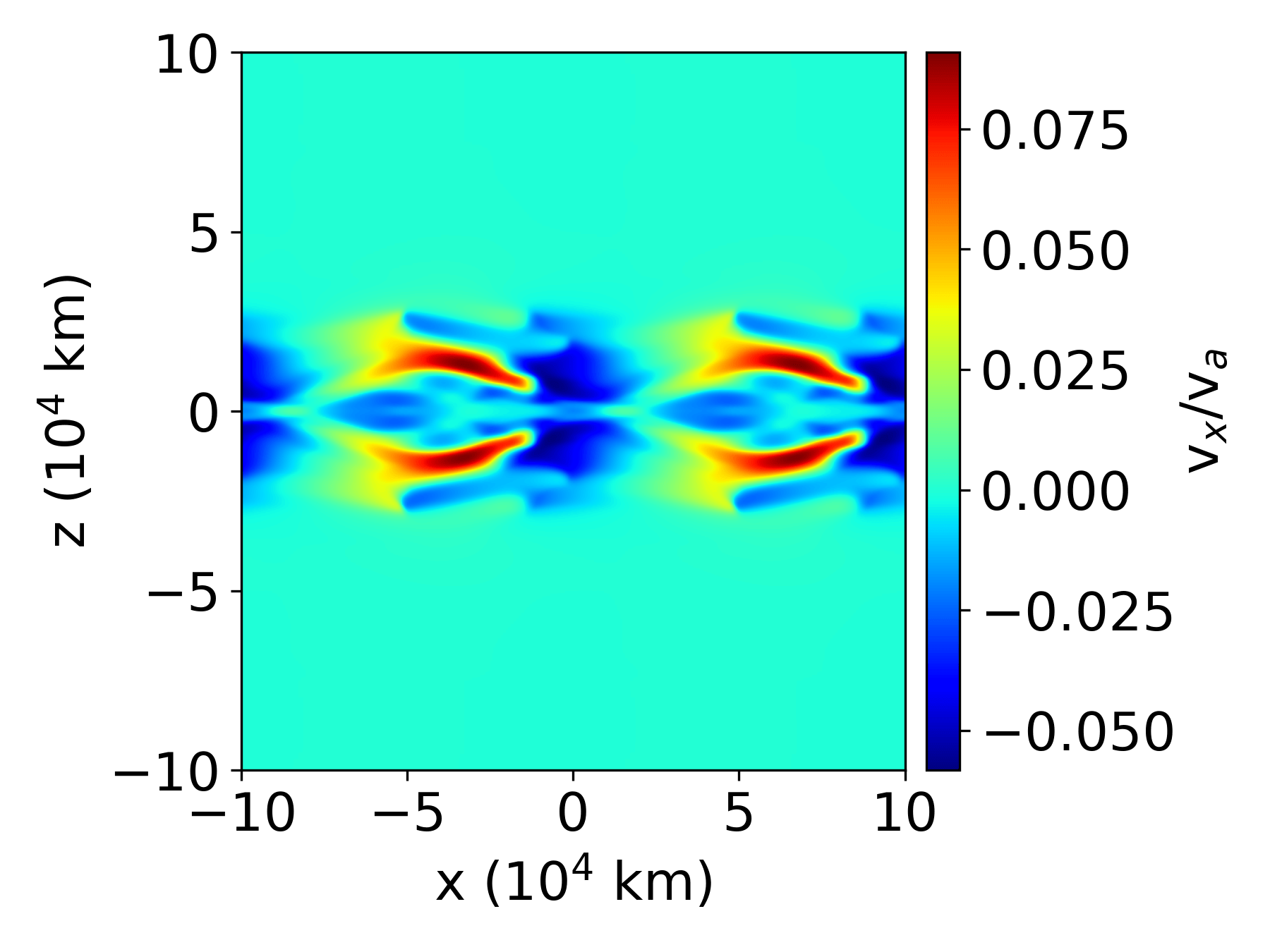}
    \caption{at $y=0$}
    \label{fig:vx_xz_t145}
\end{subfigure}
\hspace{0.5 cm}
\begin{subfigure}{0.30\textwidth}
    \includegraphics[width=1.1\textwidth]{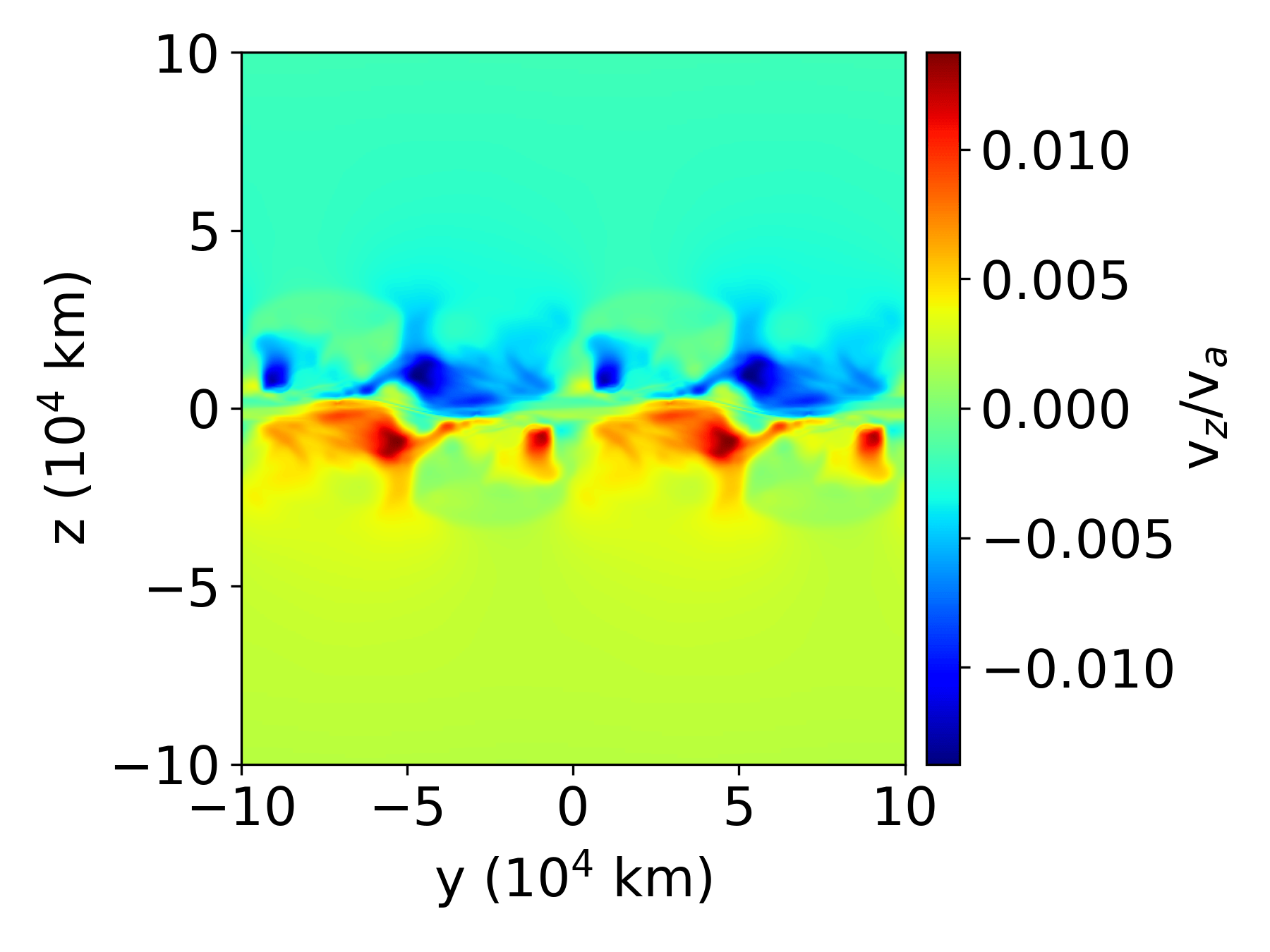}
    \caption{at $x=0$}
    \label{fig:vz_yz_t145}
\end{subfigure}
\caption{Distribution of plasma pressure (top panels) and velocity (bottom panels) at $t=207.5$ min in different planes, where the velocities scaled in units of Alfv{\'e}n velocity, $v_a \approx 261$ km s$^{-1}$.}
\label{fig:p-v-slices}
\end{figure*}

\begin{figure*}[hbt!]
\centering
\begin{subfigure}{0.3\textwidth}
    \includegraphics[width=1.1\textwidth]{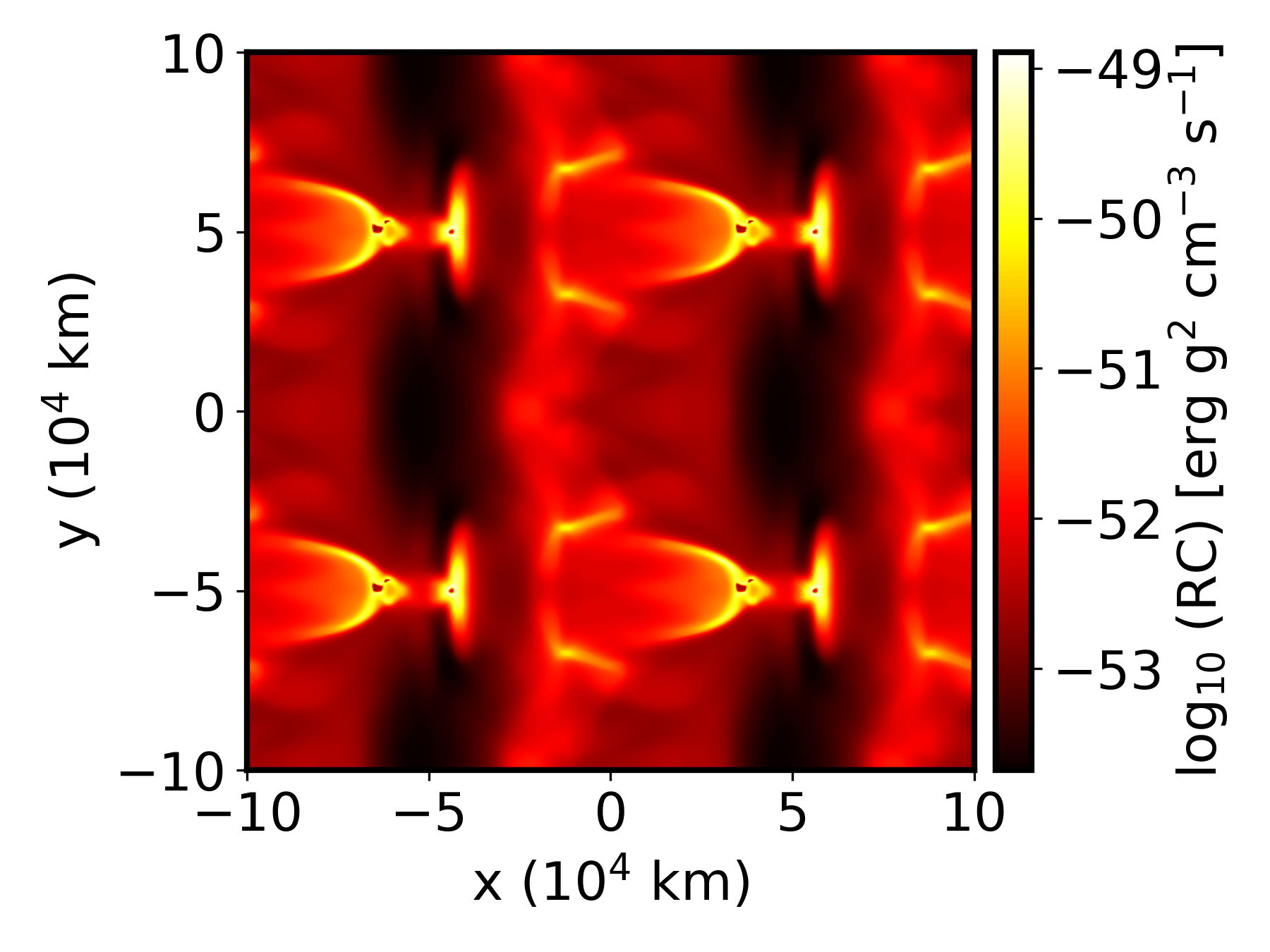}
    \caption{at $z=0$}
    \label{fig:rc_xy_t145}
\end{subfigure}
\hspace{0.5 cm}
\begin{subfigure}{0.3\textwidth}
    \includegraphics[width=1.1\textwidth]{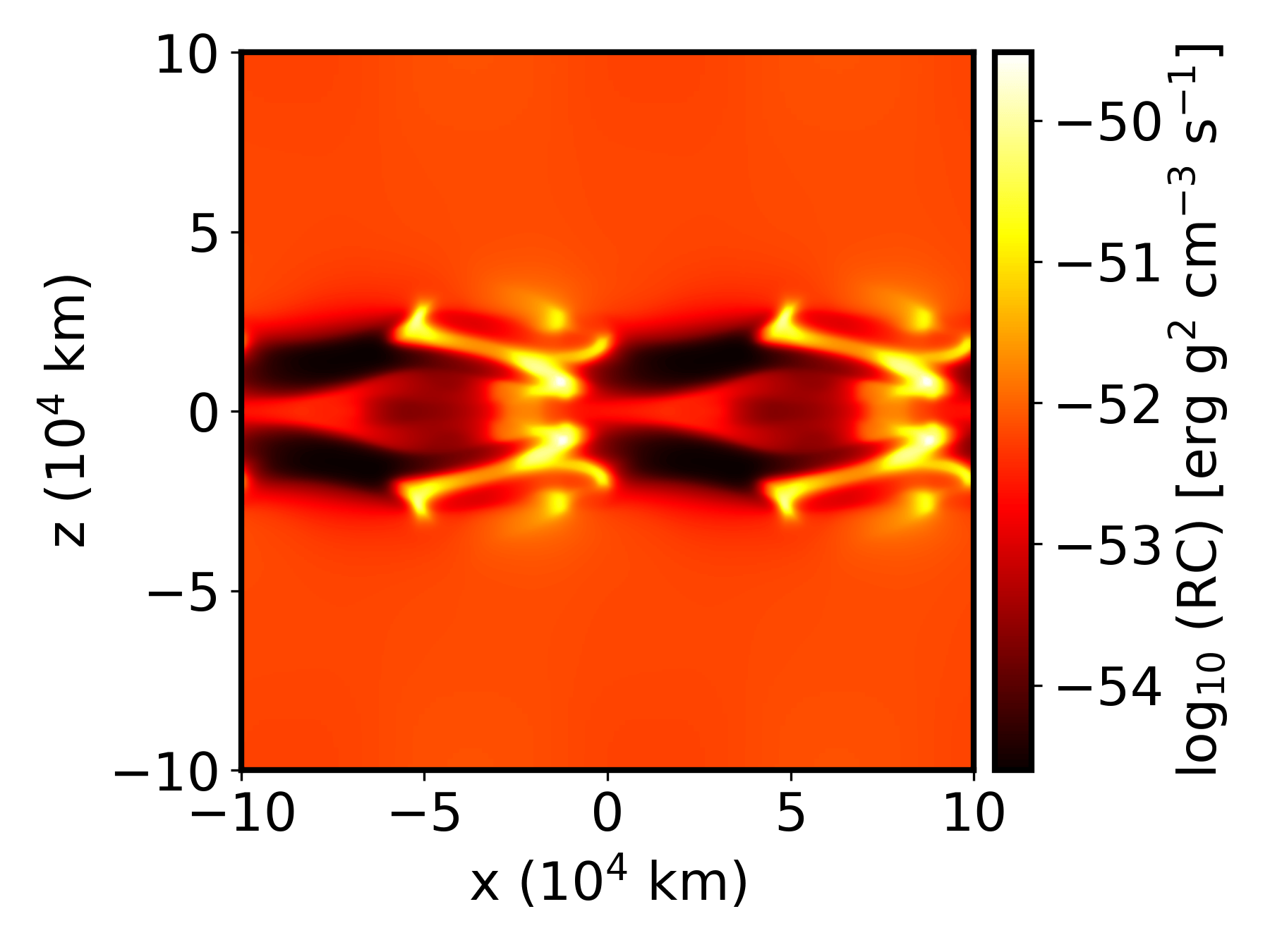}
    \caption{at $y=0$}
    \label{fig:rc_xz_t145}
\end{subfigure}
\hspace{0.5 cm}
\begin{subfigure}{0.3\textwidth}
    \includegraphics[width=1.1\textwidth]{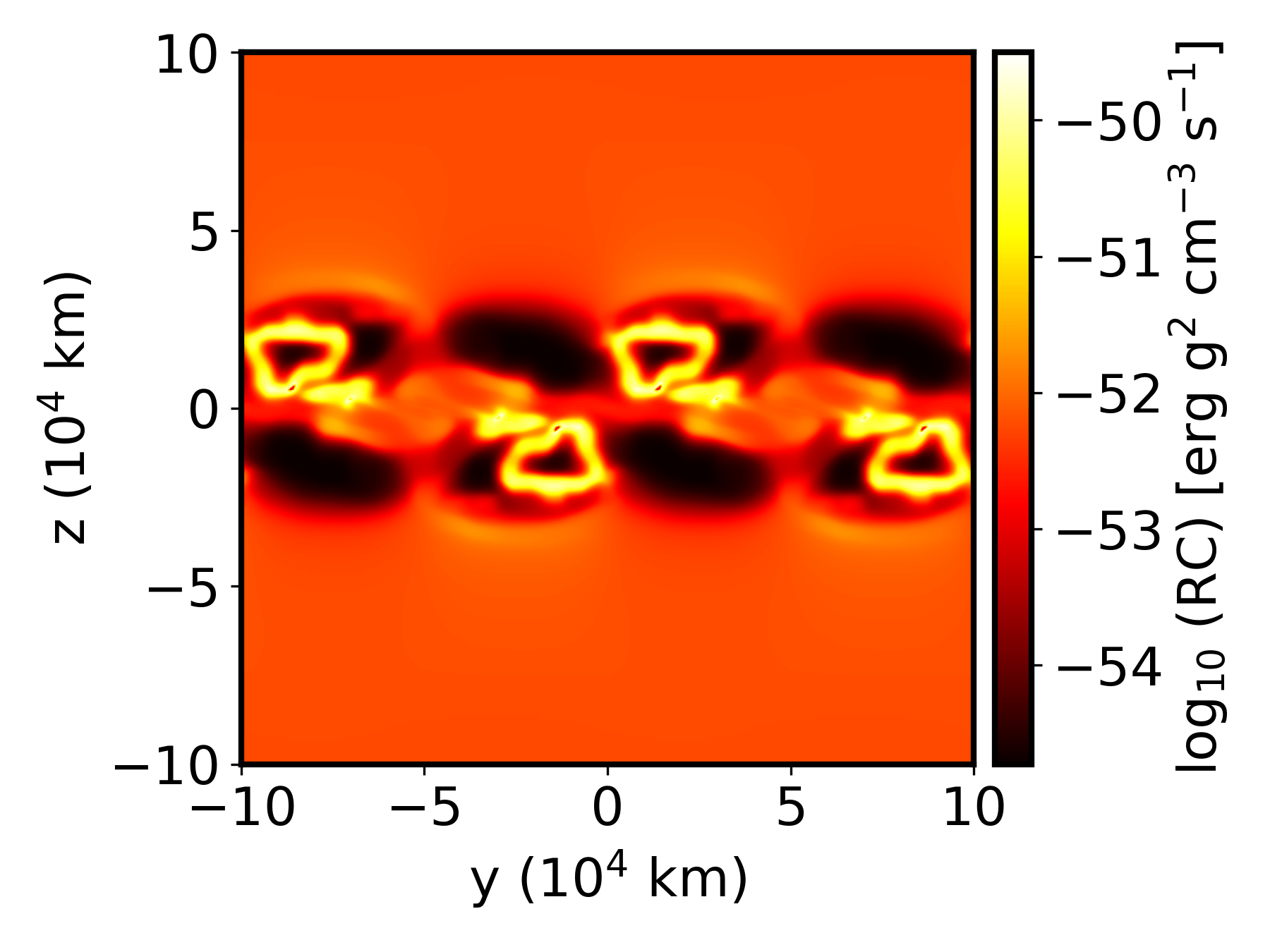}
    \caption{at $x=0$}
    \label{fig:rc_yz_t145}
\end{subfigure}
\caption{Distribution of radiative cooling (RC) losses for optically thin coronal conditions at $t=207.5$ min in different planes.}
\label{fig:rc-slices}
\end{figure*}

\begin{figure}
    \centering
    \includegraphics[width=0.4 \textwidth]{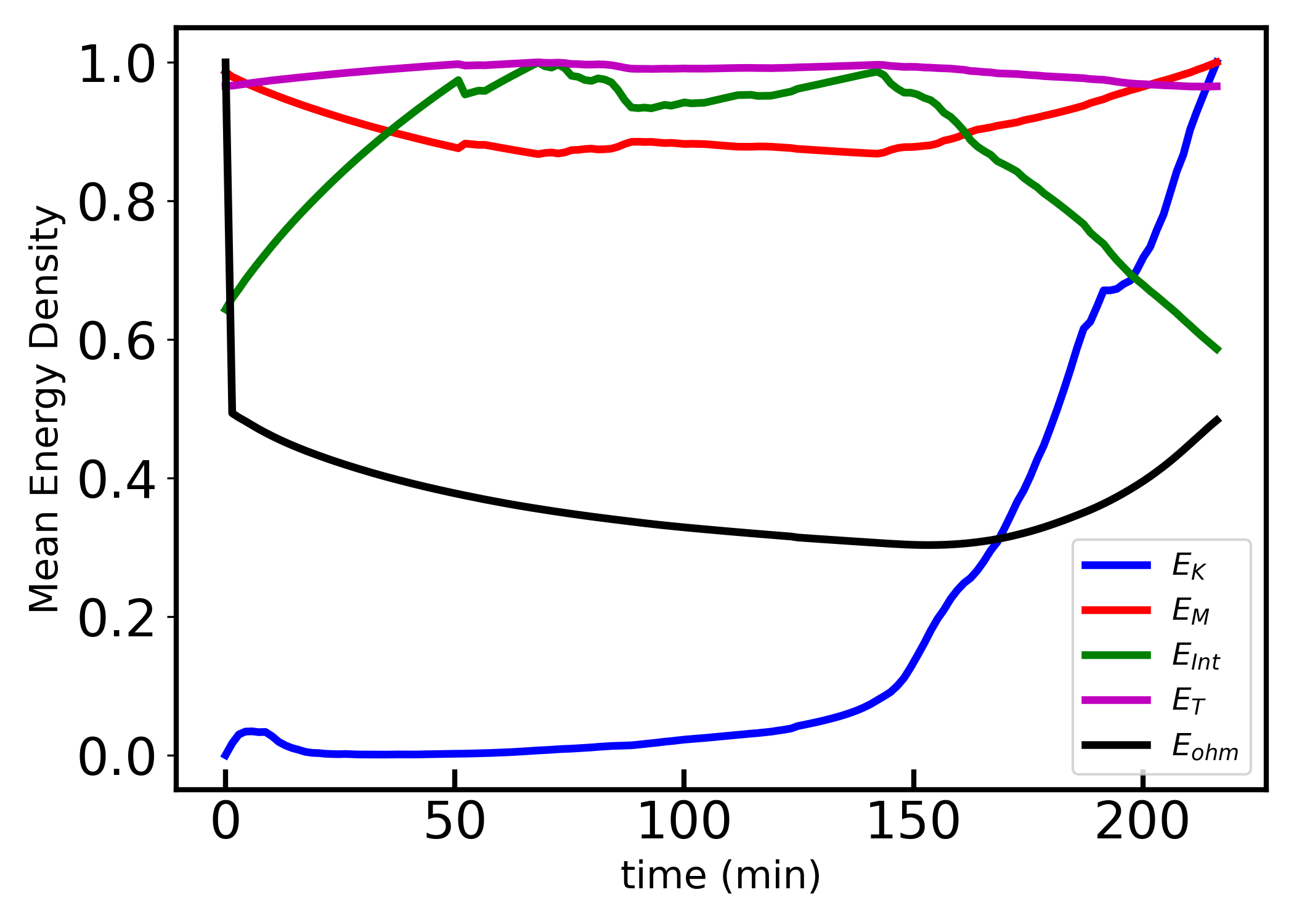}
    \caption{Time series of mean kinetic ($E_k$), magnetic ($E_M$), internal ($E_{int}$), total ($E_T$) energy density, and ohmic heating rate ($E_{ohm}$), normalised with respect to their maximum values which are $1.89 \times 10^{-3}$, $1.61 \times 10^{-1}$, $7.40 \times 10^{-2}$, $2.13 \times 10^{-1}$ erg cm$^{-3}$, and $2.23 \times 10^{-9}$ erg cm$^{-3}$ s$^{-1}$, respectively.}
    \label{fig:energydens_ts}
\end{figure}

The temporal variation of the instantaneous maximal plasma density and minimal temperature are shown in Fig. \ref{fig:den_tem_time}. The peak density increases from the initial uniform $4.68 \times 10^{-16}$ g cm$^{-3}$ at $t=0$ to $7.11 \times 10^{-14}$ g cm$^{-3}$ at $t=214.7$ min. A sharp rise of the peak density starts from $\approx 150$ min. The instantaneous minimum temperature starts from 0.5 MK (the initial uniform equilibrium temperature), which also drops sharply at the same time where the density peak has the sharp rise ($t \approx 150$ min), and reaches down to $10, 148$ K at $t=214.7$ min. This signals the formation of cool plasma condensations in the system at $\approx 150$ min, or more than two-and-a-half hours following the initial tearing onset. The variation of the instantaneous maximal velocities ($v_x$, $v_y$ and $v_z$) in Fig.~\ref{fig:peak_vel} demonstrates that a dynamical instability of the system occurs at the condensation onset time ($\approx 150$ min). Here, the velocities are scaled in terms of Alfv{\'e}n velocity, $v_a=261$ km s$^{-1}$, which is calculated based on the initial equilibrium density and the magnetic field strength of the system. We notice from Figure \ref{fig:tearing_later} that $B_z$ field evolves up to $\approx 0.4$ G along the $x-z$ plane (at $y=0$) at $t=207.5$ min, which is around an order of magnitude higher than the $B_z(x,z)$ field at $t=14.3$ min shown in Figure \ref{fig:bz_xz_t010}. The kinks appear in the bottom panel of Figure \ref{fig:tearing_later} in the $B_z$ field around $z=0$ plane (at $y=0$) along $x=\pm 2$ Mm  shows the evolution of tearing mode around the current sheet plane. The evolution of density and temperature distributions along three orthogonal planes at $x=0$, $y=0$, and $z=0$, and its 3D visualization are shown in Fig. \ref{fig:den-temp-3dview}. Density and temperature inhomogeneities appear in and around the current sheet plane that are aware of the initial multimode ($n_1=4, n_2=2$) magnetic field perturbation, as shown at $t=14.3$ mins in Figs.~\ref{fig:denslice_t010} and \ref{fig:templice_t010}, respectively. Due to tearing-associated thermal instability, localized condensed structures can be clearly seen in the $x=0$, $y=0$, and $z=0$ plane at $t=207.5$ min (see Fig. \ref{fig:denslice_t145}). The 3D visualization of the density distribution is shown in Fig.~\ref{fig:den3d_t145}, where we see the condensed structures are formed around the current sheet plane. These condensed structures correspond to the cooler ($\sim 10^4$ K) regions compared to the background medium (see Figs. \ref{fig:templice_t145} and \ref{fig:tem3d_t145}). Their order 100 density and temperature contrasts are similar to coronal rain or prominence features. Note that they happen near the evolving current sheet, which is heated up to several million degrees due to effective Ohmic heating. The periodic boundary treatments that we use in this study are acceptable for the entire evolution, since the plasmoid sizes do not reach the lateral domain size of the simulation box. The condensations from thermal instability develop locally, and are expected not to be influenced by the type of boundary used laterally. Histograms of the mass and temperature distributions in the entire domain at $t=207.5$ min are shown in Fig.~\ref{fig:mass-temp-dist}, where we see in Fig.~\ref{fig:mass_dist_t145} that 98.8\% of the cells within the simulation box contain mass within the range of $4.84 \times 10^4$ to $1.43 \times 10^5$ kg (the total number of cells in the simulation box used here is the effective resolution $512^3$), while the number of cells with mass $\gtrsim 1.43 \times 10^5$ kg is very low ($\approx 1.2 \%$). The mass range with the most number of cells is in accord with the mass determined by the initial equilibrium density of the medium (since density remains almost unperturbed away from the current sheet plane). Similarly, most cells contain the temperature value in the range of $\approx 0.32 - 0.63$ MK, namely 75.1\% of the total cells in the box as shown in Fig.~\ref{fig:temp_dist_t145}. Note that the initial equilibrium 0.5 MK temperature of the system, lies within this range. The fraction of the cell numbers with temperature $\lesssim 10^5$ K, and $\gtrsim 1$ MK are low, and they are 2.1\% and 1.2\% respectively. This implies that the cool-condensed structures, as well as the hot regions with $\gtrsim$ MK temperatures (which appear around the current sheet plane due to reconnection-induced heating) are very localised in the medium. To appreciate the thermodynamics of the cool condensed structures, we show slices of plasma pressure and different velocity components in Fig.~\ref{fig:p-v-slices}. We notice that the velocities that develop in the different cutting planes in Figs.~\ref{fig:vx_xy_t145}-\ref{fig:vz_yz_t145} are associated with pressure gradient driven flows (see Figs.~\ref{fig:p_xy_t145}-\ref{fig:p_yz_t145}), also called siphon flows. They are directed from higher to lower pressure regions and demonstrate sub-Afv{\'e}nic speeds (Alfv{\'e}nic Mach number reaches up to $\approx$ 0.075). Due to plasma accumulation as evident from the velocity maps, the cool condensation sites develop in these same regions as shown in Fig.~\ref{fig:den-temp-3dview}. When the thermal runaway sets in after a long term evolutionary process, the velocities are dominated by pressure gradient flows. In this stage initial condensation seeds merge along each other to form larger condensation sites, and drag the magnetic field lines along, which entangle and form flux ropes. The heating/cooling (mis)balance in different planes are shown in Fig.~\ref{fig:rc-slices}. We use a uniform (and constant in time) background heating of $6.244 \times 10^{-53}$ erg g$^2$ cm$^{-3}$ s$^{-1}$ which is equal to the radiative loss (no field aligned thermal conduction at the initial state due to isothermal condition) at the initial equilibrium state of the system. But this balance breaks down due to tearing influenced thermal changes around the current sheet, and the radiative loss in some regions near the current sheet plane dominates over the heating. These regions correspond to the cool condensed structures as shown in Fig.~\ref{fig:den-temp-3dview}. The regions away from the current sheet plane maintain the heating/cooling balance as the initial perturbation was concentrated around the current sheet plane, and therefore those regions maintain the initial equilibrium density and temperature of the system. The energetic evolution of the system is shown in Fig. \ref{fig:energydens_ts}. As expected (despite the open boundaries along $z$), this shows that the mean total energy density ($E_T$), which is a sum of mean kinetic ($E_k$), magnetic ($E_M$), and internal ($E_{int}$) energy densities (see Eqn.~\ref{eq:energy_density}) given by
\begin{align}
    E_k &= \frac{1}{V}\iiint_V \frac{\rho v^2}{2} {\rm d}x {\rm d}y {\rm d}z,\\
    E_M &= \frac{1}{V}\iiint_V \frac{B^2}{2} {\rm d}x {\rm d}y {\rm d}z,\\
    E_{int} &= \frac{1}{V}\iiint_V \frac{p}{\gamma -1 } {\rm d}x {\rm d}y {\rm d}z,
\end{align}
respectively (where, $V$ is the total volume of the simulation box) is nearly conserved in time. The resistivity and thermal conduction effects do not cause any deviation from total energy conservation, and only the heating/cooling misbalance may lead to net energy losses (or gains). Due to the resistive MHD evolution of the system, the energy exchange between magnetic and internal energies show an anti-correlation nature while the system evolves. This energy exchange occurs due to the mean Ohmic dissipation, which is quantified as
\begin{align}
    E_{ohm} &= \frac{1}{V}\iiint_V \eta J^2 {\rm d}x {\rm d}y {\rm d}z.
\end{align}
We notice that the mean magnetic energy density decreases up to $\approx 150$ min due to release of magnetic energy through reconnection. Thereafter, when the thermal runaway process happens in the system in a coupled tearing-thermal fashion, the field lines start to entangle with each other and form flux ropes. This generates magnetic stress, and leads to the enhancement of the magnetic energy density. However, the open boundaries along $z-$direction, which are sufficiently away from the central current sheet plane allow the magnetic energy flux to flow through the boundaries. The current density (spatially) distributes rapidly due to the implemented perturbation, and therefore we see a sharp drop of volume averaged $J^2$ initially. The temporal evolution of the mean kinetic energy density shows that the instability dynamics of the system shows a (nearly) quasi-equilibrium nature up to $\approx 150$ min (which is the onset time of condensations), and then rises sharply due to a tearing-thermal coupled unstable evolution.  

\begin{figure*}[hbt!]
    \centering
    \includegraphics[width=1\textwidth]{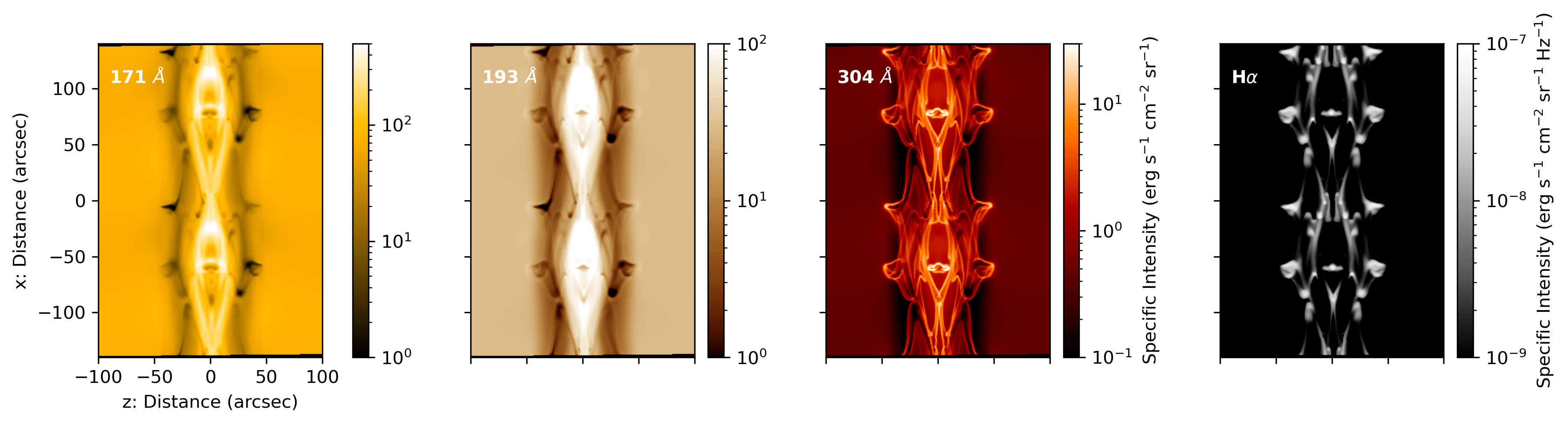}
    \caption{Specific intensity counterparts to the simulation, where we account for emission and absorption. Synthetic maps for the broadband 171, 193, 304~\AA\ SDO/AIA filters, and a narrowband Hydrogen-H$\alpha$ filter, for a LOS view along the $y$ direction at $t=207.5$~min, are shown from the left to right panels, respectively. An animation of this figure for a rotating LOS view around the $z$ (from 0 to 360$^{\circ}$) axis is available online. This shows the limb view on a current sheet that shows thermal-tearing evolutions.}
    \label{fig:synmap}
\end{figure*}

\subsection{Synthetic Observation}\label{synthetic_observation}

The \ac{SDO} is a near-Earth orbiting satellite suite capable of routinely observing the Sun from its photosphere to corona. The \ac{AIA} on board captures images of the solar atmosphere with a temporal cadence of $\sim$~12~s at a spatial resolution of $\sim$~0.6$^{''}$ per pixel across a range of \ac{UV} and \ac{EUV} wavelengths, the latter mainly associated with the different ionisation states of Iron, namely Fe~\textsc{xii}\,--\,\textsc{xxiv}. Emission from such highly ionised Iron corresponds to coronal temperatures in the broad range of a few hundred kK to around $20$~MK. The emission coefficient for such coronal plasma is given by,
\begin{equation}
    j_\lambda(\tau) = \frac{A_b}{4\pi}\,n_{e}^2(\tau)\,G_{\lambda}(n_{e}(\tau),T(\tau))\,, \label{eq:emission_coefficient}
\end{equation}
where $A_b$ is the abundance of the emitting species, $n_e$ is the ambient electron number density, and $G_{\lambda}$ is the contribution function for a specific wavelength, indicated to be additionally dependent on $n_e$ as well as temperature $T$. In the absence of a modelled $n_e$, it is instead approximated using \ac{LTE} Saha-Boltzman. This contribution function is precomputed for each of the \ac{AIA} passbands using the \texttt{CHIANTI} atomic package for a range of electron number densities and temperatures between 10$^{6}$\,--\,10$^{12}$~cm$^{-3}$ and 10$^4$\,--\,10$^8$~K, respectively \citep[][]{Landi:2013, Verner:1996}. $\tau$ here denotes the local optical depth, computed within each voxel that a given \ac{LOS} intersects as the product of the local absorption coefficient $\alpha_\lambda$ and the length of the ray within that voxel. The atmosphere of the Sun is optically thin at the wavelengths corresponding to the emission by these Fe lines. As such, the standard approach to synthesizing simulations so as to resemble the appearance of structures as seen by \ac{SDO}/\ac{AIA} is to employ Eq.~\ref{eq:emission_coefficient} in a local manner and apply an arbitrary \ac{LOS} integration according to the position of an observer. For structures that are majority-comprised of material at coronal temperatures, such as coronal loops, this is deemed a sufficient approach \citep[][]{vanDoorsselaere:2016,Gibson:2016}.

Cool condensations within the solar corona appear dim in \ac{EUV} contrast \citep[cf.][]{Carlyle:2014a}, and indeed the value of $G_{\lambda}$ for the \ac{AIA} passbands that we consider is many orders of magnitude lower at condensation temperatures than for coronal temperatures. However, the strong contrast is not only due to small $G_{\lambda}$, but also the direct absorption and removal of background EUV photons from the light beam \cite[][]{Kucera:1998}. This is due to a number of the wavelengths observed by \ac{SDO}/\ac{AIA} lying below the head of the Hydrogen Lyman continuum at 912~\AA, and so H, He, and He~\textsc{i} (with characteristic temperatures $< 10$~kK) are photo-ionised by this \ac{EUV} emission up to the ionisation continuum of He~\textsc{ii} at 227~\AA\ \citep[][]{Williams:2013}. Hence, \ac{EUV} photons are progressively removed from the \ac{LOS} if such cool material is encountered. The absorption coefficient as a consequence of this \ac{EUV} photo-ionisation can be approximated in \ac{LTE} by,
\begin{equation}
    \alpha_{\lambda} = (n_H(\tau) + n_{He}(\tau))\sum_s w_s(\tau)\,A_{b,s}\,\sigma_s,
\end{equation}
where $s$ refers to the photo-ionised element and $A_{b,s}$ and $\sigma_s$ are the
assumed abundance and cross-section of ionization of element $s$, measured observationally and experimentally/theoretically, respectively. The summation weights $w_s(\tau)$ are the ratios of the number densities as $w_{H} = 1 - n_{H\textsc{i}}/n_{H}$, $w_{He} = 1 - (n_{He\textsc{i}} - n_{He\textsc{ii}})/n_{He}$, and $w_{He\textsc{i}} = n_{He\textsc{ii}}/n_{He}$. To obtain an approximation to these weights, and in accordance with our previous assumptions of \ac{LTE}, we iteratively solve for each of the considered population densities using the Saha equation and associated partition functions. We consider convergence under the \ac{LTE} assumption to be achieved once the absolute relative difference of $n_e$ between iterations drops below an arbitrary value of $10^{-4}$ \citep[a method developed by][]{Zhou:2019}. One must necessarily consider this photo-ionisation to correctly approximate the appearance of cold plasma condensations, if present, when synthesizing simulations of the solar atmosphere \citep[][]{Jenkins:2022}.

Both the emission and absorption quantities as defined above are purely local properties, the total emergent intensity $I_\lambda(\tau_\lambda)$ along a given \ac{LOS} through these local voxels is then given by the integral form of the transport equation,
\begin{equation}
I_\lambda\left(\tau_\lambda\right)=I_\lambda(0)\,\mathrm{e}^{-\tau_\lambda}+\int_0^{\tau_\lambda} S_\lambda\left(\tau_\lambda^{\prime}\right) \mathrm{e}^{-\left(\tau_\lambda-\tau_\lambda^{\prime}\right)} \mathrm{d} \tau_\lambda^{\prime}, \label{eq:transportEq}
\end{equation}
where the combined influence of $j_\lambda$ and $\alpha_\lambda$ are taken into account in the source function $S_\lambda = j_\lambda/\alpha_\lambda$, and $\tau_\lambda$ now represents the total optical thickness along the chosen \ac{LOS}, and $I_\lambda(0)$ is the intensity of any background illumination, when the emergent intensity is calculated along the specific LOS with zero optical thickness region \citep[][]{Rybicki:1986}. The non-standard inclusion of the absorption coefficient requires every local voxel, with their respective optical depths, $\tau_\lambda'$ to have access to a globally integrated and \ac{LOS}-specific optical depth, $\tau_\lambda$. Such a requirement is not compatible with the block-based architecture of \texttt{MPI-AMRVAC} and is hence completed in post processing using a combination of \texttt{yt-project}, \texttt{numpy}, \texttt{scipy}, and \texttt{matplotlib} in python. The implementation here represents an update of that previously presented in \citet{Jenkins:2022}.

Ground-based observatories do not have access to \ac{EUV} wavelengths as recorded by \ac{SDO}/\ac{AIA}, and instead commonly observe, amongst others, the strong Hydrogen $n=3~\rightarrow~n=2$ (H$\alpha$) line at 6563~\AA. This line is known to straddle the optically-thick\,--\,thin divide under solar atmospheric conditions, and a complete handling of the plasma-light interaction of such photons through the simulation domain would require \ac{NLTE} modelling, outside the scope of this study \citep[but we point an interested reader to the recent work of ][for comparison]{Jenkins:2023}. Instead, \citet[][]{Heinzel:2015} reported an approximate approach to relating local pressure and temperature conditions to H$\alpha$ opacity ($\alpha_\lambda$) according to their series of 1.5D radiative transfer models. A key property of these models considers the source function of Eq.~\ref{eq:transportEq} to remain constant along a given \ac{LOS}, and enables the following simplification,
\begin{equation}
    I_\lambda (\tau_\lambda) = I_\lambda(0)\,e^{-\tau_\lambda} + S_\lambda(1 - e^{-\tau_\lambda}). \label{eq:constStransferEq}
\end{equation}
The resulting emergent (specific) intensity of the H$\alpha$ line is therefore found through a \ac{LOS} integration of the approximate $\alpha_\lambda$ according to the tables of \citet{Heinzel:2015}, wherein the authors also provide a coarse height-dependent estimation to the constant value of $S_\lambda$.

To convert the physical variables (plasma density and temperature) into spectroscopic observables (namely specific intensity), we generate the synthetic maps of the simulation output using forward modeling. Measuring LOS integrated specific intensity depends on the (theoretical) viewing position of the observer. Thus, due to the spatial distribution of the substructures due to the condensations, the synthetic maps show differences for different LOS views. Fig. \ref{fig:synmap} shows the synthetic maps of the reoriented spatial domain (we now show the current sheet vertically) capturing the internal structures as being viewed sideways, representative for the solar limb. We do so for three different \ac{EUV} passband filters of AIA, 171, 193, and 304 \AA\, using the contribution functions of their specific spectral lines, which each highlight material at $\approx 0.8$ MK, 1.5 MK, and 80 kK respectively, for a LOS integration along our $y$ direction at $t=207.5$~min. An animation of the synthetic maps for different LOS directions around the $z$ axis is available online. Due to the absorption features of the condensed plasma regions, the strongest absorption corresponds to the LOS direction along the $y$ direction, as the condensations are aligned with it (see Fig.~ \ref{fig:den-temp-3dview}). The cool, condensed plasma appears darker in AIA 171, 193, and 304 \AA\ passband filters due to photo-ionisation of H, He, and He~{\sc i} as previously outlined. For the remaining optical H$\alpha$ line, we obtain a positive intensity in the absence of any background illumination, as would be the case for, amongst others, prominences and jets positioned at or above the limb. The island like structures near $z=0$ in the EUV maps in Fig.~\ref{fig:synmap} represent plasmoids, which are the manifestation of the extended flux ropes along $y$ direction as shown Fig.~\ref{fig:fl_t145}. The central current sheet is hotter than the surroundings, and therefore we see the widened area of the bright band in AIA 171, and 193 filters. The dark small features that appear in the EUV maps are due to absorption from the dense materials as those are located along the $y$-direction. These cool materials with temperature $\sim 10^4$ K appear bright in H$\alpha$ map. The synthetic observation for other LOS directions (shown in the animation) demonstrate a wide range of cool dense structures distributed along $x$ and $y$ directions.    

The resolution of the presented simulation is 390 km$^2$ per pixel. In all cases, the spatial resolution of our synthesized images have been modified to match that of an equivalent observatory. For the EUV pass bands, this is set to the instrumental resolution of the AIA filters ($\approx$ 430 km$^{2}$ per pixel), and leads to a minor smearing of the finer structures inside the finest condensations. This blurring is stronger for the H$\alpha$ line where the resolution is set to 1$^{''} = 725.27$~km so as to match the GONG ground-based instruments \cite[][]{Harvey:2011}. It will be useful to increase our model resolution for comparisons against the state-of-the-art observations anticipated from Solar Orbiter High Resolution Imager (HRI) campaigns \citep{2020A&A...642A...8R}.

\section{Summary and Conclusions}\label{summary_conclusions}

The study of the combined tearing-thermal instability of an idealized 3D current sheet configuration as addressed in this work is to understand the theoretical basis of the multi-mode evolution of a current sheet, which is an important aspect to understand the dynamics and multi-thermal behaviour of solar atmosphere. In contrast to our earlier 2D simulations of a non-force-free current sheet (SK22), where thermal runaway was happening simultaneously with chaotic tearing, and condensations were trapped and gathered within coalescing plasmoid structures, the current setup shows a clear tearing evolution at first leading to 3D topological changes in the magnetic field, and later on show runaway condensations near the central current sheet. Note that there are some important differences between the initial setup of the current model with the one used in SK22, as explained in section \ref{numerical_setup}, most notably perhaps the plasma beta regime which is uniformly low in the current 3D simulation. Here, the condensation onset time in the current model is much later than in SK22, because the initial setup and adopted perturbations do not directly modify the heat-loss balance and a purely tearing evolution starts at first. In a 3D long-term simulation of a macroscopic current sheet, we find that cool plasma condensations are produced in the vicinity of the current sheet due to tearing-influenced thermal instability. Our findings are based on a 3D resistive MHD simulation with non-adiabatic effects of radiative cooling (for an optically thin medium), background heating, and magnetic field-aligned thermal conduction, which are relevant for the solar corona. We find that the plasma density of the condensations can go up to $\sim10^{-14}$ g cm$^{-3}$, which is two orders of magnitude more than the initial background density ($4.6 \times 10^{-16}$ g cm$^{-3}$), and the temperature of the condensations can drop down to $\sim 10^4$ K, which is an order of magnitude less than the initial equilibrium temperature (0.5 MK) of the medium. These locally, in-situ forming cool condensed structures hence show similar thermodynamic contrasts with their surroundings, just as coronal rain or prominences observed in the solar atmosphere. This is highlighted by synthetic views, which account for important absorption effects within the synthesis of the EUV channels of \ac{SDO}/\ac{AIA}. However, our model ignores the effect of the stratified solar atmosphere due to solar gravity. The condensation time scales in this idealized current sheet model is $\approx 150$ min, which is larger than the time scales (which is $\approx 30$ min) compared to the earlier study for a post-flare coronal rain model by \cite{2021:wenzhi}, where they use a stratified atmosphere, and reveal the multi-thermal aspects of a post flare loop underneath the current sheet and reconnection sites. Whereas, our study demonstrates the development of multi-thermal plasma ($\sim 10$ kK - MK) in and around the current sheet and reconnection sites. Also, earlier models of coronal rain in arcades show a clear tendency to develop strong shearing motions \citep{2022ApJ...926..216L, 2021:yuhao, 2015:fang}, and velocity shear can alter the tearing mode growth, which we plan to explore by incorporating velocity shear flows into the model.  Future efforts should exploit a more realistic 3D model by using the initial magnetic field configuration from an extrapolated magnetic field for an active region vector magnetogram. Nevertheless, the current study sheds new light on the instability of solar current sheets due to the combined effect of tearing and thermal modes, and unifies multi-thermal processes in current sheets with the formation mechanism of cool condensations such as prominences, and coronal rain in solar atmosphere, which are important aspects in the understanding of the broader solar coronal heating.

\begin{acknowledgements}
Data visualization and analysis are performed using \href{https://visit-dav.github.io/visit-website/index.html}{Visit} and \href{https://yt-project.org/}{yt-project}. SS and RK acknowledge support by the C1 project TRACESpace funded by KU Leuven. JMJ and RK acknowledge the support by the European Research Council (ERC) under the European Unions Horizon 2020 research and innovation program (grant agreement No. 833251 PROMINENT ERC-ADG 2018) and a FWO project G0B4521N. The computational resources and services used in this work were provided by the VSC (Flemish Supercomputer Center), funded by the Research Foundation Flanders (FWO) and the Flemish Government - department EWI. RK acknowledges the International Space Science Institute (ISSI) in Bern, ISSI international team project \#545.
\end{acknowledgements}

 \bibliographystyle{aa} 
 \bibliography{ref, bibliography} 

\begin{appendix}
\section{Evolution of the system with $T_0=1$ MK temperature}\label{app}
\begin{figure}[hbt!]
\centering
\begin{subfigure}{0.3\textwidth}
    \includegraphics[width=1.1\textwidth]{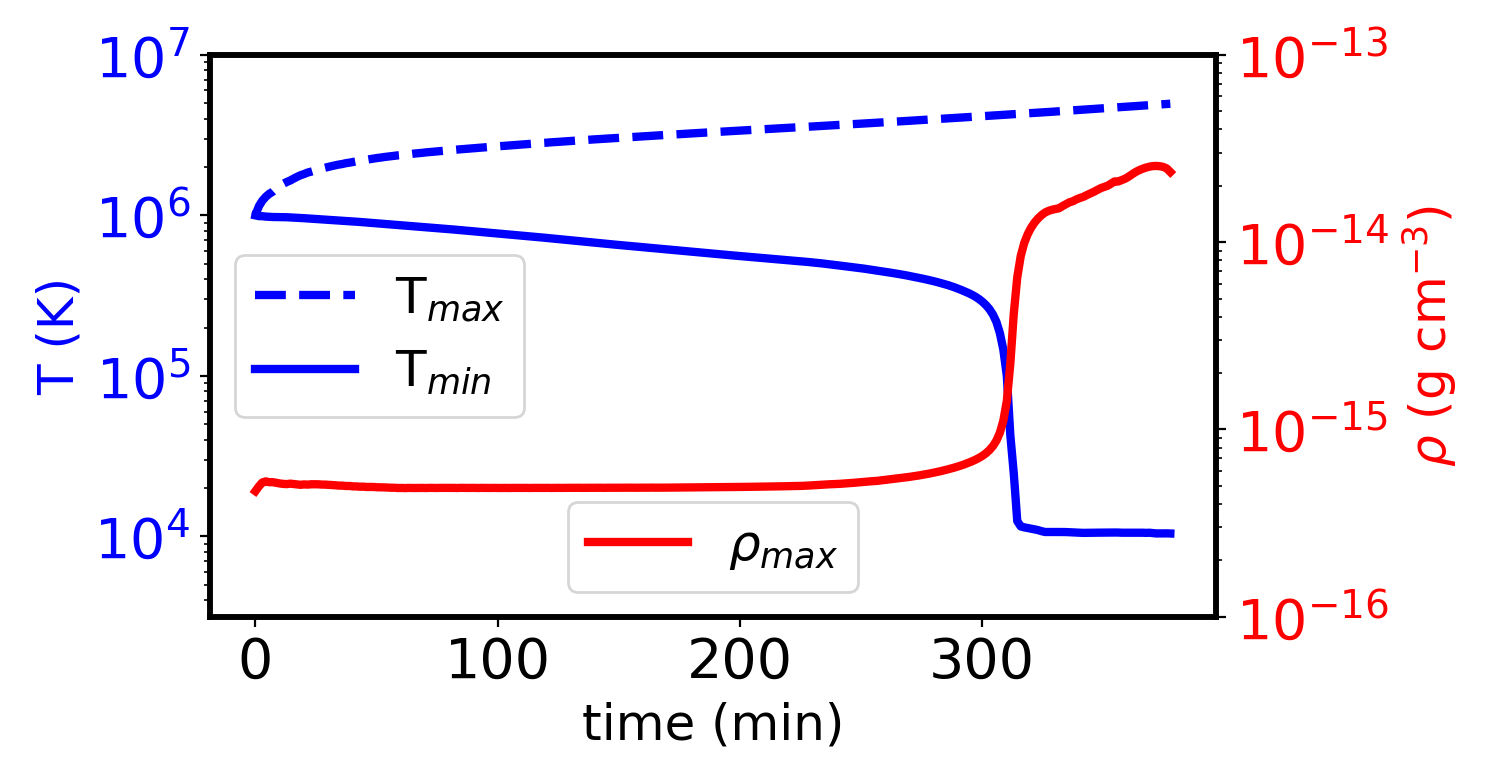}
    \caption{}
    \label{fig:app1a}
\end{subfigure}
\begin{subfigure}{0.35\textwidth}
    \includegraphics[width=1\textwidth]{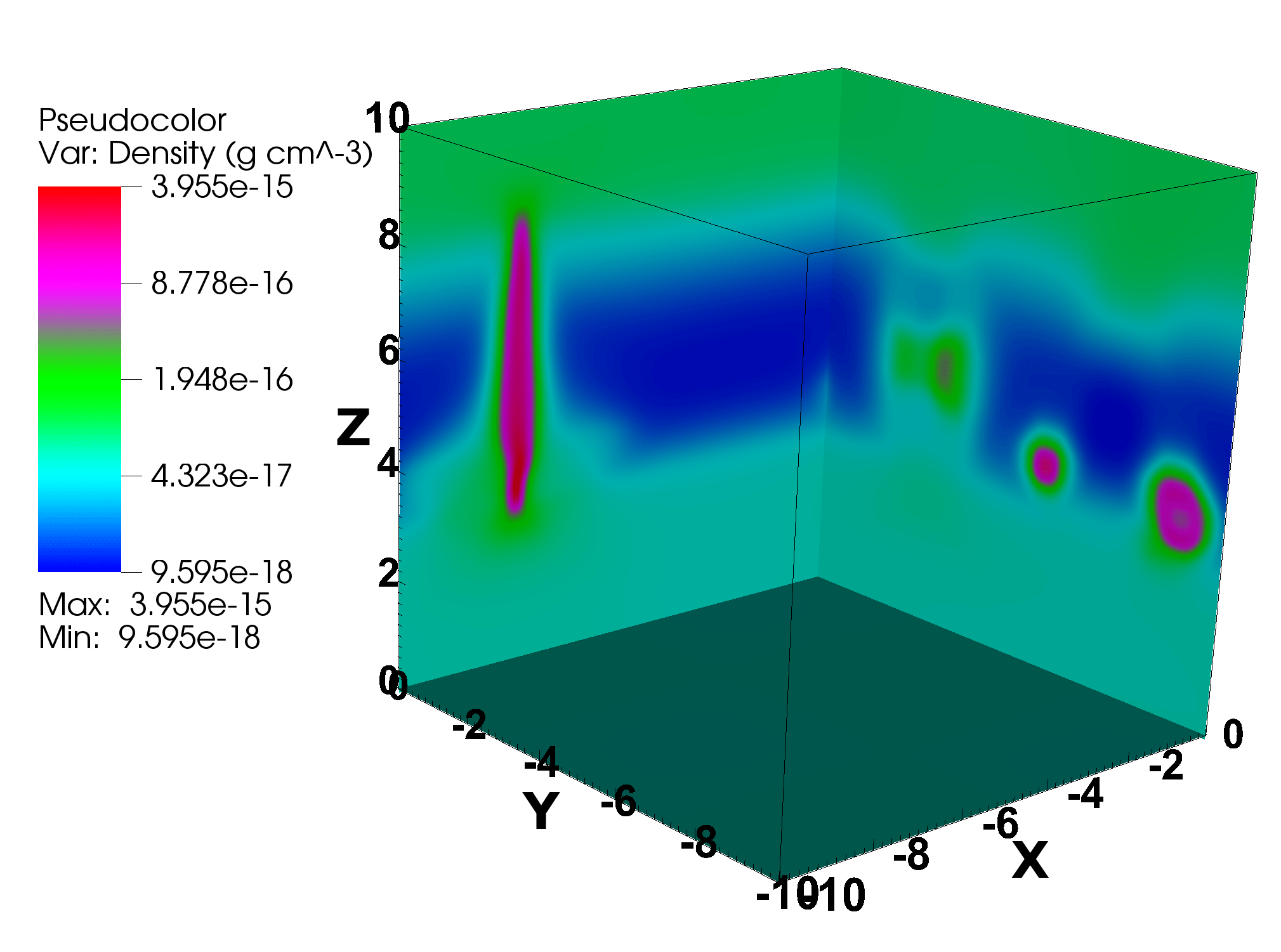}
    \caption{}
    \label{fig:app1b}
\end{subfigure}
\begin{subfigure}{0.35\textwidth}
    \includegraphics[width=1\textwidth]{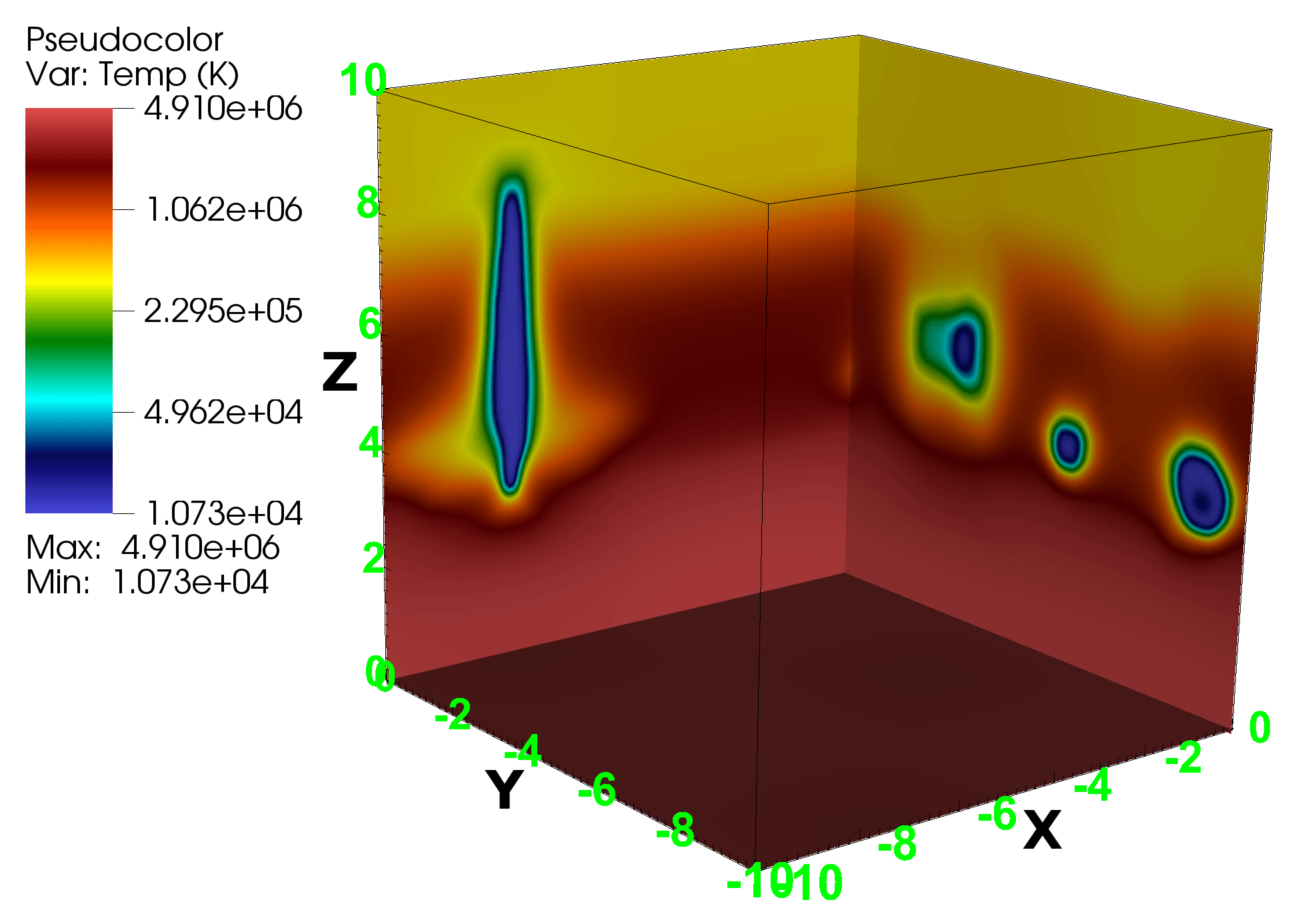}
    \caption{}
    \label{fig:app1c}
\end{subfigure}
\caption{(a) represent the temporal evolution of instantaneous peak density (red curve), and maximum (blue dashed), and minimum (blue solid) temperature. (b) and (c) represent the density and temperature distribution between $x, y \in [-10, 0]$, and $z \in [0, 10]$ at $t=372$ min respectively, where $x, y$, and $z$ are in units of $10^4$ km.}
\label{fig:app1}
\end{figure}

To investigate whether the condensation occurs in the experiment with a different equilibrium temperature, we run the simulation with $T_0=1$ MK keeping all other parameters same. In this case we use a lower resolution with AMR level up to 2, which gives the maximum resolution of $256^3$ along $x$, $y$, and $z$ directions. Figure \ref{fig:app1a} represents that the instantaneous peak density increases more than an order of magnitude at $\approx 300$ min, and the minimal temperature drops down to $\sim10$ kK (which is 2 orders of magnitude less from the initial temperature) at the same point of time, whereas the maximal temperature rises from 1 MK to $\approx5$ MK. From Figures \ref{fig:app1b} and \ref{fig:app1c}, we notice that the cool ($\sim 10^4$ K) and condensed ($\sim 10^{-15}$ g cm$^{-3}$) structures are spatially co-located, which demonstrates that the condensation occurs in our experiment is a common phenomena in different temperature ranges, and not specific to a particular temperature regime.

\end{appendix}

\end{document}